\documentclass[english,useAMS,usenatbib]{mnras}
\usepackage{bm}
\usepackage[T1]{fontenc}
\usepackage[latin9]{inputenc}
\usepackage{units}
\usepackage{rotating}
\usepackage{url}
\usepackage{amsmath}
\usepackage{commath}
\usepackage{amssymb}
\usepackage{graphicx}
\usepackage[caption=false]{subfig}
\usepackage{pdfpages}
\usepackage{esint}
\usepackage[authoryear]{natbib}
\usepackage{listings}
\usepackage{textcomp}
\usepackage[export]{adjustbox}
\usepackage[para,online,flushleft]{threeparttable}

\usepackage{etoolbox}
\makeatletter
\patchcmd\@combinedblfloats{\box\@outputbox}{\unvbox\@outputbox}{}{%
   \errmessage{\noexpand\@combinedblfloats could not be patched}%
}%
 \makeatother

\def\gtsima{$\; \buildrel > \over \sim \;$}
\def\ltsima{$\; \buildrel < \over \sim \;$}
\def\prosima{$\; \buildrel \propto \over \sim \;$}
\def\gsim{\lower.5ex\hbox{\gtsima}}
\def\lsim{\lower.5ex\hbox{\ltsima}}
\def\simgt{\lower.5ex\hbox{\gtsima}}
\def\simlt{\lower.5ex\hbox{\ltsima}}
\def\simpr{\lower.5ex\hbox{\prosima}}
\def\la{\lsim}
\def\ga{\gsim}

\newcommand{\be}{\begin{eqnarray}}
\newcommand{\ee}{\end{eqnarray}}
\def\lsim{\,\lower2truept\hbox{${< \atop\hbox{\raise4truept\hbox{$\sim$}}}$}\,}
\def\gsim{\,\lower2truept\hbox{${> \atop\hbox{\raise4truept\hbox{$\sim$}}}$}\,}

\providecommand{\tabularnewline}{\\}

\title[Evolution of local dwarfs as GW150914 hosts]{Evolution  of dwarf galaxies hosting GW150914-like events}
\author[Marassi et al.]{S. Marassi$^{1,2}$\thanks{E-mail:
stefania.marassi@roma1.infn.it} and L. Graziani$^{1,2}$\thanks{E-mail:
Luca.Graziani@roma1.infn.it}, M. Ginolfi$^{1,2}$, R. Schneider$^{1}$, M. Mapelli$^{3,4,5}$ \newauthor 
M. Spera$^{3,4,5}$, M. Alparone$^{1,6}$\\ 
$^{1}$Dipartimento di Fisica, Sapienza, Universit$\grave{a}$ di Roma, Piazzale Aldo Moro 5, 00185, Roma, Italy\\
$^{2}$INAF / Osservatorio Astronomico di Roma, Via di Frascati 33, 00078 Monte Porzio Catone, Italy\\
$^{3}$INAF/Osservatorio Astronomico di Padova, Vicolo dell' Osservatorio 5, I-35122, Padova, Italy \\
$^{4}$INFN, Sezione di Milano-Bicocca, Piazza della Scienza 3, 20126 Milano, Italy \\
$^{5}$Institut f\"ur Astro- und Teilchenphysik, Universit\"at Innsbruck, Technikerstrasse 25/8, A--6020, Innsbruck, Austria\\
$^{6}$Dipartimento di Ingegneria, Universit\`a degli Studi di Napoli Parthenope, Centro Direzionale Isola C4, 80143 Napoli, Italy\\}
\begin{document}

\date{Maggio 2017}

\pagerange{\pageref{firstpage}--\pageref{lastpage}} \pubyear{2016}

\maketitle

\label{firstpage}

\begin{abstract}

Here we present a detailed analysis of the properties and evolution of different dwarf galaxies, candidate to 
host the coalescence of black hole binary systems (BHB) generating GW150914-like events. By adopting a novel theoretical 
framework coupling the binary population synthesis code \texttt{SeBa} with the Galaxy formation model \texttt{GAMESH}, 
we can investigate the detailed evolution of these objects in a well resolved cosmological volume of 4~cMpc, having a 
Milky Way-like (MW) galaxy forming at its center. We identify three classes of interesting candidate galaxies: MW progenitors, dwarf satellites and dwarf galaxies evolving in isolation. We find that: (i) despite differences in individual histories and specific environments the candidates reduce to only nine representative galaxies; (ii) among them,  $\sim44\%$ merges into the MW halo progenitors by the redshift of the expected signal, while the remaining dwarfs are found as isolated or as satellites of the MW and their evolution is strongly shaped by both peculiar dynamical history and environmental feedback; (iii) a stringent condition for the environments where GW150914-like binaries can form comes from a combination of the accretion history of their DM halos and the radiative feedback in the high redshift universe; (iv) by comparing with the observed catalogues from DGS and ALLSMOG surveys we find two observed dwarfs respecting the properties predicted by our model. 
We finally note how the present analysis opens the possibility to build future strategies for host galaxy identification.

\end{abstract}


\section{Introduction}
With the first  detection of the two signals GW150914  and GW151226 \citep{2016PhRvL.116f1102A, 2016PhRvL.116x1103A} 
the LIGO Collaboration opened a new era of gravitational astronomy relying on gravitational wave (GW) observations with  high signal to noise ratios (SNR): 24 and 13, respectively. 
These GWs, interpreted as the result of two coalescing black holes (BH), 
have now two similar companion events detected during the second run when Advanced Virgo \citep{2015CQGra..32b4001A} joined the LIGO detectors: GW170104 ($\rm SNR=13$,  \citealt{2017PhRvL.118v1101A}) and GW170814 ($\rm SNR=18$, \citealt{2017PhRvL.119n1101A})\footnote{Note that during the second run the lightest black hole binary so far observed was also detected (GW170608, \citealt{2017ApJ...851L..35A})  with components in the range inferred for low-mass X-ray binaries: $\rm M_1 \sim 12$~$M_\odot$ and $\rm M_2 \sim 7$~$M_\odot$, respectively.}. 

The masses inferred for the components of the BHB generating GW150914 are $\rm M_1 \sim 36$~$M_\odot$ and $\rm M_2 \sim 29$~$M_\odot$; BHBs associated with GW170104 (and GW170814) are massive as well, with inferred component masses of $\rm M_1 \sim 31$~$M_\odot$ ($\sim 30$~$M_\odot$), $\rm M_2 \sim 19$~$M_\odot$ ($\sim 25 $~$M_\odot$), respectively\footnote{  During the referee process of our paper the LIGO-Virgo Collaboration announced the detection of four coalescing massive black hole binaries, GW170729, GW170809, GW170818, GW170823, we comment this recent result in Section \ref{sec:Discussion}.}. 

These massive BH binary systems are believed to generate from progenitor stars with low metallicity \citep{2010ApJ...714.1217B,2013ApJ...779...72D,2016Natur.534..512B,2016MNRAS.459.3432M,2016MNRAS.463L..31L,2016ApJ...818L..22A,2017PhRvL.118v1101A,2017ApJ...850L...4C,2017MNRAS.464.2831O,2017MNRAS.471L.105S,2017NatCo...814906S,2018MNRAS.474.2959G}. Stellar models predict in fact that only low-metallicity progenitors can generate these massive remnants due to their weak stellar winds, and consequently reduced mass loss \citep{2008NewAR..52..419V,2009MNRAS.395L..71M,2010ApJ...714.1217B,2015MNRAS.451.4086S,2017MNRAS.470.4739S}. 

The metallicity of the stars, combined with the redshift of the signal and an estimate of the binary coalescence time, can also constrain the galactic environment in which these massive BHBs are allowed to form. For example, if GW150914 had  merged in a short time, it could have formed in a low-metallicity, low redshift dwarf galaxy \citep{2016ApJ...818L..22A,2016Natur.534..512B}. Conversely, a longer merging time would have allowed its formation in a wider class of galaxies at high redshift, where low-metallicity environments are more common \citep{2016MNRAS.463L..31L,2017MNRAS.464.2831O,2017MNRAS.471L.105S,2017MNRAS.472.2422M}. Also note that the detected signals do not provide sufficient constraints on alternative scenarios on the formation of these binaries, either dynamical or isolated (\citealt{2016ApJ...818L..22A} and references therein). Future detections of black hole spins could help disentangling them, as discussed in \citet{2010CQGra..27k4007M,2017Natur.548..426F,2017MNRAS.471.2801S}. Finally,  
it is worth mentioning that GW150914 and GW170104 could be explained also by alternative scenarios as highlighted in \citet{2016PhRvL.117f1101S,2017arXiv170604211D,2018arXiv180205273B}, and references therein.

After the first detection, a series of studies investigated conceivable formation  
environments of these massive binaries and their merger rates by adopting different approaches (see for example \citet{2016Natur.534..512B,2016MNRAS.461.3877D}). Recent models based on galaxy scaling relations \citep{2016MNRAS.463L..31L,2018MNRAS.473.1186E} suggest a low redshift BHB merger rate 
as a function of the present-day host galaxy mass, while \cite{2017MNRAS.464.2831O} predict a BHB merger 
rate in dwarf and massive galaxies by adopting cosmological simulations of galaxies evolving down to $z=0$. 

Models combining BHBs population-synthesis codes with cosmological simulations have been introduced by many authors by post-processing cosmological simulations. \cite{2017MNRAS.472.2422M} post-processed the Illustris cosmological simulation \citep{2014MNRAS.444.1518V}  with the results of novel population-synthesis simulations performed with the code MOBSE \citep{2018MNRAS.474.2959G}, to investigate the cosmic history of BHB mergers on a cosmological scale of $\sim 100$~cMpc. They find that most of the  GW150914-like systems that merge in the local Universe formed at high redshift with a long delay time. \cite{2018arXiv180103099L} coupled the same population synthesis model with a high-resolution hydro-dynamical zoom-in simulation of a MW-like galaxy \citep{2016ApJ...827L..23W} to provide an estimate of the number of BHBs mergers. 

A hybrid theoretical framework to study the formation and coalescence sites of BHBs in a self-consistent cosmological evolution was introduced in \citet{2017MNRAS.471L.105S} (hereafter SC17). This approach is based on the coupling between the \texttt{GAMESH} pipeline (\citealt{2015MNRAS.449.3137G}, GR15) and the population synthesis code {\texttt{SeBa}
\citep{1996A&A...309..179P,2001A&A...375..890N,2013MNRAS.429.2298M,2014ApJ...794....7M}, and allows to perform a Monte Carlo 
simulation of BHB formation on top of a high-resolution dark matter (DM) run. In this way is it possible to describe the evolution of DM halos and their hosted galaxies in a highly resolved cosmic volume of 4~cMpc. 
This volume is selected to contain  a Milky Way-like halo in its center and a statistically significant population of companion and satellite galaxies (mostly of dwarf type). The simulation evolves from $z\sim 20$, down to the Local Universe ($z=0$) through a complex interplay of mechanical, chemical and radiative feedback (the interested reader can find more details in the next section or in the original references: GR15, \citet{2017MNRAS.469.1101G} (GR17), \citealt{2018MNRAS.473.4538G}). Thanks to the high mass resolution of the embedded N-Body simulation, \texttt{GAMESH} can follow the baryonic evolution of  structures down to a minimum mass of $\sim 3.4\times10^7$~M$_{\odot}$. These galaxies are the best candidate to host events of star formation in low-metallicity environments (hence massive BH formation).
 
SC17 find that $\sim 50\%$ of GW150914 and $\sim60\%$ of GW151226 binary systems are hosted in galaxies that are MW progenitors at the event redshifts. These have masses of $\rm M_{\star}\sim 4\times 10^{10}M_{\odot}$, metallicity  $\rm Z \sim 0.4 Z_{\odot}$ and form stars at rates of $\rm \sim 5 M_{\odot}/yr$. 
Moreover, $10\%$  of GW150914-like systems are found in very small galaxies, with $\rm M_{\star}\sim 5\times 10^{6}M_{\odot}$ where 
star formation is suppressed by feedback effects. 
 
In the present work we investigate the evolution of representative galaxies hosting GW150914-like events, also found in SC17 and provide details on their baryonic and dynamical properties as well as on their evolutionary pathways. We finally search for galaxies with similar properties in observational samples (see Section \ref{sec:sample}). 

The plan of the paper is as follows. In Section 2 we briefly describe  the galaxy formation simulation, while Section 3 describes the observational data-set we adopted to compare simulated dwarfs with observed ones. The properties of galaxies hosting the birth and the coalescence of BHB systems associated with  GW150914-like events are discussed in Section 4. In particular, by following the merger tree of \texttt{GAMESH}, we can describe both the dynamical and baryonic evolution of the galaxies in which GW150914-like binaries live. A comparison with two observed dwarfs is then made. Finally, we discuss the properties of the most interesting evolutionary channels of the selected dwarfs: they are followed from their formation redshift, down to the epoch at which the GW150914-like binary forms. Section 5 summarizes our conclusions.

\section{The galaxy formation simulation}
The galaxy formation simulation adopted in this paper is detailed in GR17 and was run with the hybrid 
pipeline \texttt{GAMESH} introduced in GR15. By interfacing a 
DM zoom-in simulation\footnote{The simulation was performed with the 
numerical scheme GCD+ \citep{2003MNRAS.340..908K, 2013MNRAS.428.1968K} and ensures a DM particle mass 
$m_p = 3.4 \times10^5$~M$_{\odot}$ in a well resolved volume of 4~cMpc 
centered in the MW-like halo. Halos are resolved by a FoF algorithm with a minimum of 100 particles and the selected volume is smaller than the global region contaminated by high-resolution particles only (a cube of about 5~cMpc/side) This choice also ensures that the dwarfs described at the borders have a reasonable stable close dynamical environments and are correctly dynamically bounded.} of an isolated MW-like halo with a 
data-constrained semi-analytic model, the latest version of 
\texttt{GAMESH} is capable to perform fast galaxy formation simulations 
accounting for the baryonic evolution of the MW and its surrounding galaxies 
under different chemical and radiative feedback schemes.

The central Milky Way halo is predicted to have a correct mass value ($M_{\texttt{MW}} \sim 1.7 \times 10^{12} M_{\odot}$), 
as well as structural properties and cosmic evolution in agreement with recent 
cosmological simulations (see  GR17 for more details).
As added value, the simulated volume provides a statistically significant number of well resolved mini- and 
Ly$\alpha$-cooling halos\footnote{Halos are usually classified on these two categories depending on their virial temperature ($T_{vir}$). if $T_{vir} \leqslant 2\times 10^4$~K the halo is considered a mini-halo, otherwise it is classified as a Ly$\alpha$-cooling halo. Note that in the  \texttt{GAMESH} feedback scheme this property is used to establish its capability to form stars by fueling gas from the surrounding environment See GR15 and references therein for a more detailed explanation of the radiative feedback scheme.} orbiting the central MW, despite the fact that the adopted 
N-Body simulation was not performed to ensure the structural and dynamical 
properties observed in the Local Group. Hereafter we will refer to this 4~cMpc 
cosmic region as the "LG" of the present simulation. The good statistic of our LG allows us to explore 
the dynamical evolution of a large sample of mini-halos ($\sim 2800$) surviving the MW potential at $z=0$ 
through dynamical interactions. In addition, many Ly$\alpha$-cooling 
halos having DM masses compatible with structures hosting intermediate mass galaxies (as the observed  M32, M33 or LMC-type galaxies) are present and can be studied with a high level of detail.

As the star formation in \texttt{GAMESH} is calibrated to ensure that the
MW galaxy has the observed values of stellar, gas and metal mass (hereafter M$_{\star}$, 
M$_{\rm gas}$, M$_{Z}$) at $z=0$, the stellar populations of the main galaxy 
can be described with a high level of accuracy and their history explored in detail. 
Furthermore, the properties of the surrounding structures satisfy a large number of 
observational constrains: the observed galaxy main sequence, the mass-metallicity 
and the fundamental plane of metallicity relations in $0 < z < 4$. Moreover, \texttt{GAMESH} accounts 
for the correct stellar mass evolution of candidate MW progenitors in $0 \lesssim z \lesssim 2.5$, 
although their star formation rate and gas fraction follow a shallower redshift dependence (GR17 and Appendix A).

The coupling with  the binary population synthesis code \texttt{SeBa} (in its modified version 
by \citealt{2013MNRAS.429.2298M,2014ApJ...794....7M}) is described in SC17 and Appendix B. \texttt{SeBa} simulates different channels of  binary system formation in 11 bins of gas metallicity, sampling the range $\rm 0.01Z_{\odot}\leq Z\leq 1Z_{\odot}$.   For each simulation, $N = 2\times 10^{6}$ binaries are explored, having initial physical properties randomly selected from independent distribution functions (for more details on the simulated samples adopted in this study see Appendix B). Thanks to the particle-based merger-tree of \texttt{GAMESH}, sites of BHB formation and coalescence can be univocally linked and followed through their cosmic evolution. For example, halos found in the coalescence redshift range predicted by the LIGO GW150914 detection can be followed back in time and their history reconstructed through dynamical events (i.e. mergers/tidal interactions/halo disruptions) and baryonic evolution of the galaxies (e.g. star formation/metal pollution). The formation of binaries along the cosmic evolution is accounted for by applying a certain binary fraction to the newly formed stellar mass and by randomly sampling systems from the \texttt{SeBa} database until the mass in binaries is exhausted. As a result, many binary system types (often repeated with a certain multiplicity) are associated with each galaxy during episodes of star formation (i.e. at each star forming reshift $z_f$). Note that the newly created binaries at $z_f$ are followed in their stellar evolution consistently with galaxy assembly of their hosts under the assumption that, once formed, they just live in their halos/galaxies and evolve in their descendants following the DM merger tree. It should be noted that hydrodynamical interactions perturbing these systems within the galaxy cannot be accounted for in our model. The reader interested in more technical details is referred to Appendix C.

In this paper we consider only binaries created with masses compatible with GW150914-like events, with the additional requirement that the coalescence redshift $z_c$ is found within the LIGO detection interval. Their hosting galaxies and DM halos are selected to guarantee that at $z_f$: (i) they are star forming (i.e. SFR $>0$), (ii) in order to form GW150914 binaries their gas metallicity is sub-solar ($Z \leq 0.5 Z_{\odot}$) (iii) their newly formed stellar mass in binaries includes at least one system with the appropriate mass ratio of GW150914\footnote{ We remind here that this condition is constrained by the random sampling of BHB candidates within the {\texttt{SeBa}} catalogue until the total mass of newly formed binary systems is saturated. See SC17 and Appendix C for more details.} and finally, (iv) the lifetime of their candidate binaries allows a coalescence in the redshift range inferred by LIGO.

Finally note that each system is identified and traced as unique combination of the binary system ID (provided by \texttt{SeBa}), halo ID and evolutionary channel ID (gID, provided by the \texttt{GAMESH} merger tree).   

\section{Observational data sample}
\label{sec:sample}

Once the dwarf galaxies hosting GW150914-like binaries were found in the \texttt{GAMESH} simulation, we searched for observed galaxies with similar properties in the Dwarf Galaxy Survey (DGS; \citealp{2013PASP..125..600M}) and APEX low-redshift legacy survey of molecular gas (ALLSMOG; \citealp{2017A&A...604A..53C}). 
\newline
The DGS, extensively described in the Dwarf Galaxy Survey Overview by \cite{2013PASP..125..600M} and by \cite{2014A&A...563A..31R, 2015A&A...582A.121R}, is a sample of 48 local low-metallicity and low-stellar mass galaxies, with metallicity ranging from 12 + log(O/H) = 7.14 to 8.43 and stellar masses from $3\times10^6$ to $\sim 3 \times 10^{10}~M_\odot$. The 
DGS sample was selected from several deep optical emission line and photometric surveys including the Hamburg/SAO Survey and the First and Second Byurakan Surveys (e.g., \citealp{1991A&A...247..303I,2003A&A...397..463U}).
The gas-phase metallicity of the DGS galaxies, listed in \cite{2014A&A...563A..31R}, are provided  by empirical strong emission-line methods (\citealp{2013PASP..125..600M}), obtained through the $R_{23}$ ratio with the Pilyugin \& Thuan calibration \citep{2005ApJ...631..231P}.
The stellar masses are computed by \cite{2014PASP..126.1079M} adopting the prescription of \cite{2012AJ....143..139E} and the Spitzer/IRAC luminosities at 3.6 $\mu$m and 4.5 $\mu$m. 
The star formation rates (SFR) are computed combining two of the most widely adopted tracers: the far-ultraviolet (FUV) and the H$_\alpha$ luminosities (\citealp{2009ApJ...703.1672K}, \citealp{2012ARA&A..50..531K}).
\newline
The ALLSMOG survey comprises 88 nearby, star-forming galaxies with stellar masses in the range $10^{8.5} < M_\ast/M_\odot < 10^{10}$, and gas-phase metallicity 12 + log(O/H) > 8.4. 
This sample is entirely drawn from the MPA-JHU\footnote{\href{http://wwwmpa.mpa-garching.mpg.de/SDSS/DR7/}{http://wwwmpa.mpa-garching.mpg.de/SDSS/DR7/}} catalogue of spectral measurements and from the galactic parameters of the Sloan Digital Sky Survey, Data Release 7 (SDSS DR7; \citealp{2009ApJS..182..543A}). 
The metallicity adopted here is computed through the N2 calibration provided by Pettini \& Pagel \citep{2004MNRAS.348L..59P}).
Stellar mass and SFR values come from the MPA-JHU catalogue, where $M_{\rm{star}}$ is computed from a fit to the spectral energy distribution (SED) obtained using SDSS broad-band photometry \citep{2004MNRAS.351.1151B, 2007ApJS..173..267S}, and the SFR is based on the H$_\alpha$ intrinsic line luminosity following the method of \citep{2004MNRAS.351.1151B}. 
\newline
Together, these two samples of nearby galaxies span a wide range of physical properties ($\sim 2$ dex in metallicity, $\sim 4$ dex in stellar mass and $\sim 3.5$ in star formation rate), providing a suitable benchmark for comparisons with our simulation (see Appendix A).

\section{Results}
In this section we summarize our results. Section 4.1 describes the properties and evolution of the most representative galaxies hosting the  formation of candidate BHBs at various $z_f$. Once identified, they are followed down to their descendant galaxies at $z_c$, where these binaries coalesce generating GW150914-like signals. A comparison with observed analogues is then provided in Section 4.2. Finally, Section 4.3 investigates the early evolution of three interesting candidate dwarf galaxies predicted by our model.

\subsection{Evolution of formation and coalescence hosts}

\begin{table}
\begin{centering}
\begin{tabular}{|c|c|c|c}
\hline
gID & $z_f \rightarrow z_c$ & $\sim T_{vir}$~[$10^4$~K] & $\sim d_{\texttt{MW}}$~[100~ckpc] \tabularnewline
\hline
\hline
F1 & $3.93\rightarrow0.095$ & $2.20\rightarrow1.50$  & $24.2\rightarrow7.4$ \tabularnewline
\hline
F2  & $2.36\rightarrow0.087$ & $2.05\rightarrow7.23$  & $19.9\rightarrow4.4$ \tabularnewline
\hline
F3  & $3.55\rightarrow0.080$ & $2.20\rightarrow24.5$  & $21.5\rightarrow11.6$ \tabularnewline
\hline
E1 & $3.38\rightarrow0.072$ & $3.10\rightarrow17.66$  & $25.0\rightarrow27.4$ \tabularnewline
\hline
E2 & $4.15\rightarrow0.103$ & $2.83\rightarrow34.20$  & $24.7\rightarrow34.4$ \tabularnewline
\hline
MW1 & $2.45\rightarrow0.095$ & $2.01\rightarrow176$  & $7.2\rightarrow0.0$ \tabularnewline
\hline
MW2 & $3.93\rightarrow0.095$ & $5.43\rightarrow176$  & $11.6\rightarrow0.0$ \tabularnewline
\hline
MW3 & $2.97\rightarrow0.057$ & $3.81\rightarrow174$  & $7.1\rightarrow0.0$ \tabularnewline
\hline
MW4  & $2.54\rightarrow0.103$ & $2.18\rightarrow177$  & $10.0\rightarrow0.0$ \tabularnewline
\hline
\end{tabular}
\par\end{centering}
\caption{\label{tab:fCGalaxies} Properties of individual dark matter halos in which GW150914 BHB candidates form and coalesce, as determined by the \texttt{GAMESH} merger tree. gID is the unique evolutionary channel identifier containing both the formation and coalescence events. $z_f$ and $z_c$ are, respectively, the redshift of BHB formation and coalescence, $T_{vir}$~[K] the halo virial temperature, while $d_{\texttt{MW}}$ indicates the distance of the halo from the cube center (i.e. the MW halo) in units of 100~kpc. For each quantity the value at formation and coalescence is indicated. Also note that MW-type galaxies in the table are all progenitors of the unique MW found at the center of the highly resolved volume of 4~cMpc.}
\end{table}

Here we describe the properties of both dark matter halos and galaxies in which binary systems generating GW150914-like events form. We also connect them with their descendants by following their dynamical evolution through the \texttt{GAMESH} merger tree, until $z_c$ is reached and these galaxies can be identified as coalescence sites.  

Table 1 summarizes their DM halo properties: halo evolution path ID (gID), redshift of stellar binary formation ($z_f$) and coalescence as BHB ($z_c$), halo virial temperature ($T_{vir}$) and the distance $d_{\texttt{MW}}$ of each halo from the central MW.  Figure 1 complements this table by providing the DM mass  ($M_{h}$) of the halos at $z_f$ and $z_c$. Formation halos are shown as blue dots and connected to the coalescence ones (asterisk-like shapes) by arrows. Cyan and light-yellow shaded areas represent the redshift ranges of formation ($2.36\leq z_f \leq4.15$) and coalescence ($0.057\leq z_c \leq 0.103$), as constrained by our simulation and by the LIGO detection of GW150914 \citep{2016PhRvL.116f1102A}. 

\begin{flushleft}
\begin{figure}
\includegraphics [clip=true, width=9.50cm]{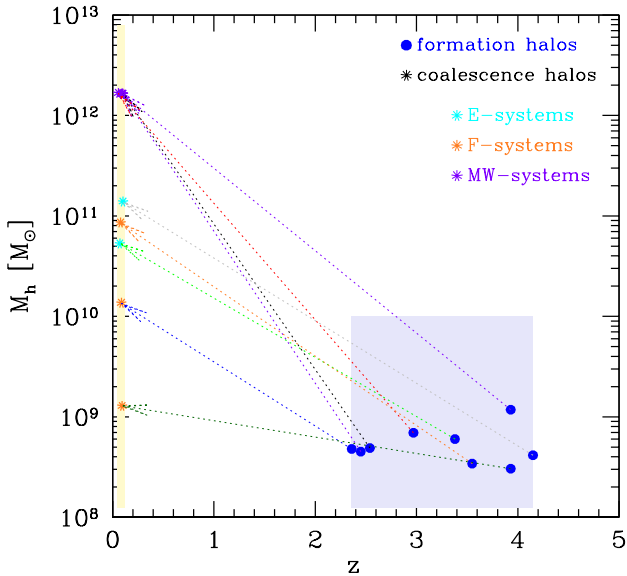}
\vspace{-1.5cm}
\caption{Dark matter mass ($M_h$) of halos hosting the formation of the stellar binary (blue dots) and 
               the BHB coalescence of GW150914 events (star-like points), as function of redshift. Cyan and light-yellow 
               shadow areas represent the formation ($2.36\leq z_f \leq4.15$) and coalescence ($0.057\leq z_c \leq 0.103$) 
               redshift ranges. Different galaxies are classified as E-, F- and MW systems depending on their 
               dynamical state. Formation halos are connected to the coalescence ones by arrows.}
\label{fig:fcHalos_DMMass}
\end{figure}
\end{flushleft}

\begin{table*}
\begin{centering}
\begin{tabular}{|c|c|c|c|c}
\hline
gID & $z_f \rightarrow z_c$ & log(M$_{\star}$)~[M$_{\odot}$] & SFR~[M$_{\odot}$/yr] & Z~[12+log(O/H)] \tabularnewline
\hline
\hline
F1 & $3.93\rightarrow0.095$ & $6.2\rightarrow6.9$  & $0.012\rightarrow0.0$ & $7.528\rightarrow7.698$\tabularnewline
\hline
F2  & $2.36\rightarrow0.087$ & $5.9\rightarrow8.3$  & $0.015\rightarrow0.016$ & $7.314\rightarrow8.119$\tabularnewline
\hline
F3  & $3.55\rightarrow0.080$ & $6.26\rightarrow9.2$  & $0.015\rightarrow0.210$ & $7.483\rightarrow8.187$\tabularnewline
\hline
E1 & $3.38\rightarrow0.072$ & $6.5\rightarrow9.2$  & $0.018\rightarrow0.105$ & $7.482\rightarrow8.333$\tabularnewline
\hline
E2 & $4.15\rightarrow0.103$ & $6.4\rightarrow9.2$  & $0.017\rightarrow0.169$ & $7.592\rightarrow8.283$\tabularnewline
\hline
MW1 & $2.45\rightarrow0.095$ & $5.9\rightarrow10.6$  & $0.0147\rightarrow5.272$ & $7.329\rightarrow8.290$\tabularnewline
\hline
MW2 & $3.93\rightarrow0.095$ & $6.7\rightarrow10.6$  & $0.061\rightarrow5.272$ & $7.533\rightarrow8.290$\tabularnewline
\hline
MW3 & $2.97\rightarrow0.057$ & $6.4\rightarrow10.7$  & $0.025\rightarrow5.078$ & $7.451\rightarrow8.294$\tabularnewline
\hline
MW4  & $2.54\rightarrow0.103$ & $6.1\rightarrow10.6$  & $0.017\rightarrow5.327$ & $7.337\rightarrow8.288$\tabularnewline
\hline
\end{tabular}
\par\end{centering}
\caption{\label{tab:barFCGalaxies} Galaxies in which the formation of binary systems and their coalescence as BHBs occur. gID is the unique galaxy evolutionary channel identifier, $z_f$ and $z_c$ are, respectively, the redshifts at which hosted BHB form and coalescence, M$_{\star}$ is the galaxy stellar mass, SFR is its star formation rate  and Z is its gas metallicity. For each quantity the galactic values at formation and coalescence of the BHBs is indicated.}
\end{table*}
Note that the number of halos where GW150914-like binaries are able to form\footnote{i.e. here we restrict the total number of GW150914 binaries created along the redshift evolution of our volume to the ones found in the redshift range predicted by LIGO, and with the right $M_1/M_2$ ratio.} is small with respect to the total number present in our cosmic volume: we find only 9 halos with mass in the range $3 \times 10^8$~M$_{\odot} \lesssim M_h \lesssim 2 \times 10^9$~M$_{\odot}$  among $\sim 130000$ total simulated halos (i.e. both dwarfs and MW progenitors) in the interesting redshift range. They are also spread  in a wide formation redshift range as a result of the long merger times predicted for massive BHB systems (3.87 - 4.2~Gyr) and of the scatter in the redshift of coalescence inferred from the data. In particular, three classes of halos hosting GW150914-like events are found in our sample: systems belonging to the main MW progenitor (MW-systems), systems belonging to MW satellites (F-systems) and halos hosting dwarf galaxies that can evolve in isolation and are not dynamically captured by the global infall onto the central Milky Way (E-systems). Note that E-systems are systematically found at the borders of our high-resolution region.

Four GW150914-like events are found in halos at $0.057 \leq z_c \leq 0.087$. Three are hosted by F2, F3 and E1 (see Table~1), while the  GW150914-like system predicted at the lowest redshift ($z_c \sim 0.057$) comes from a BHB in MW3. The remaining five sites are found in $0.095 \leq z_c \leq 0.103$; among them, two are hosted by F1 and E2 while the other signals are found in MW progenitors.  Note that MW progenitor galaxies hosts GW150914-like events with a wide range of possible formation redshift: $2.45 \leq z_f  \leq 3.93$; this naturally comes from the fact that the major branch of the MW gravitationally captures a wide ensemble of objects hosting GW150914-like binaries coalescing in the LIGO redshift range.

The small number of representative hosts described above allows us to perform a detailed inspection of their properties and evolutionary histories. Their initial virial temperatures are found to span a range $T_{vir} \sim [2.0 -5.4] \times 10^4$~K  with 55\% of the dark matter halos having values close to the lower limit and then being fragile, in our model, to both dynamical and radiative feedback (e.g. by mass stripping / star formation suppression).

\begin{flushleft}
\begin{figure*}
\hspace{-1.0cm}
\includegraphics [clip=true, width=6.76cm]{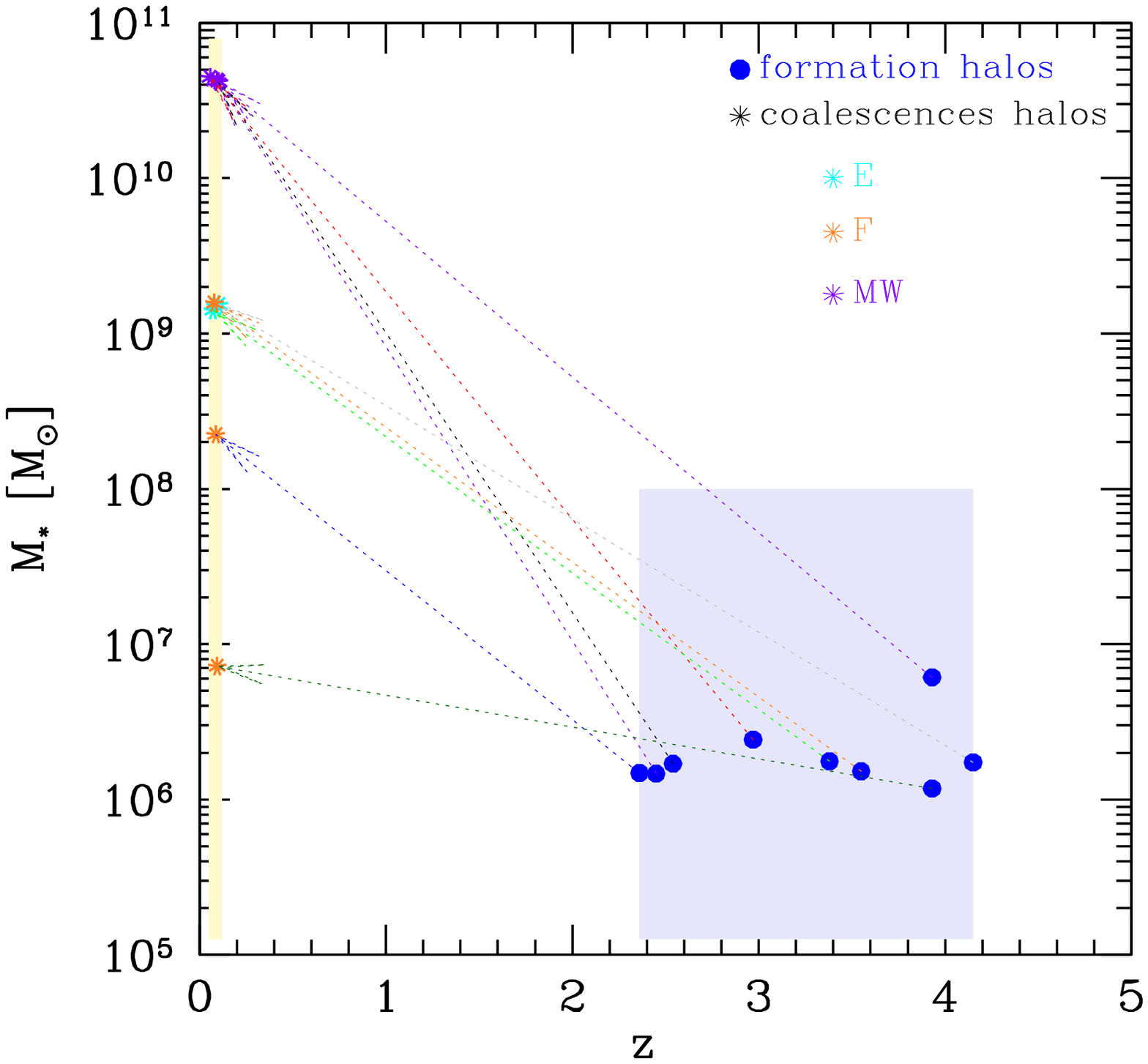}
\hspace{-0.95cm}
\includegraphics [clip=true, width=6.76cm]{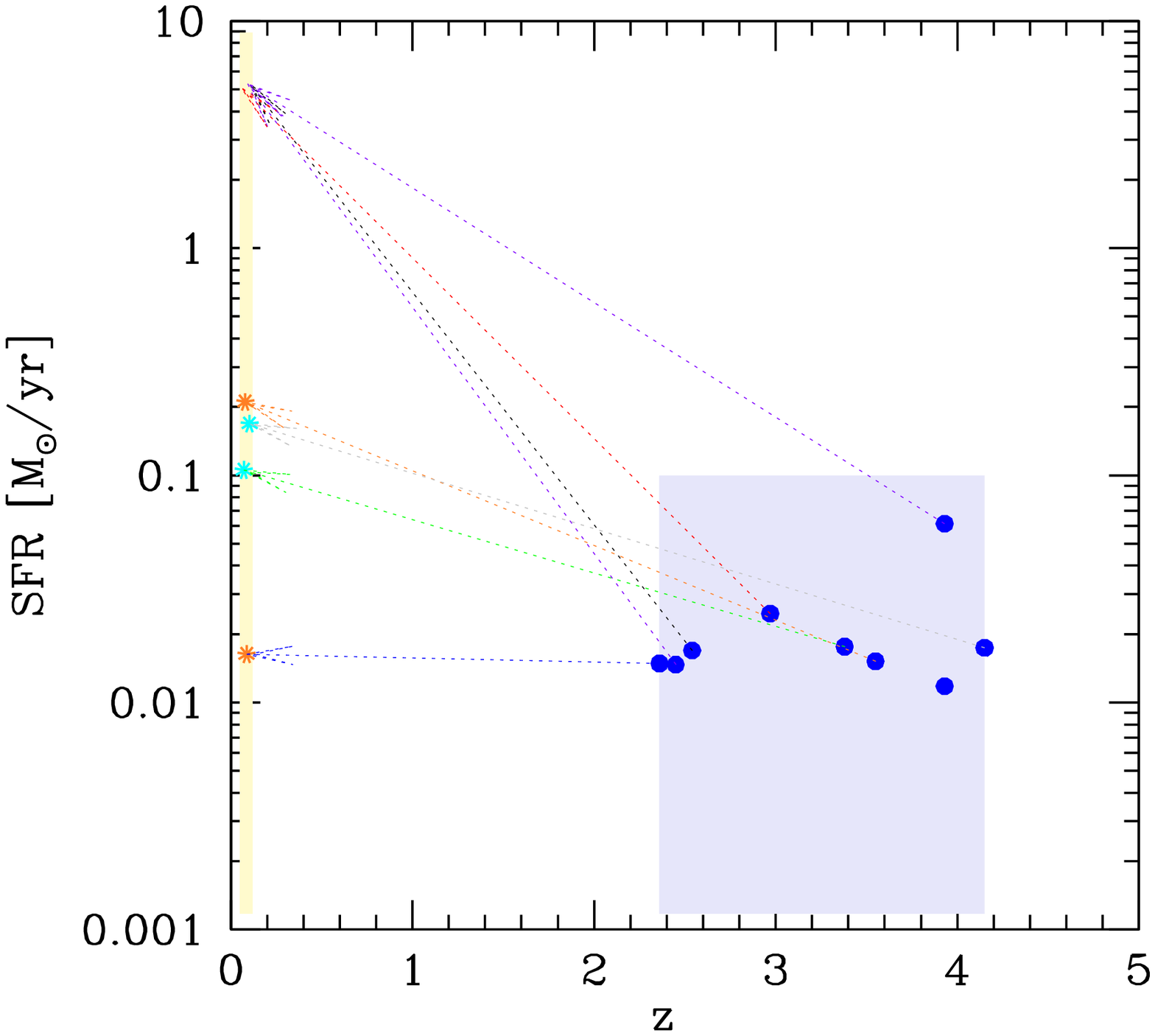}
\hspace{-1.10cm}
\includegraphics [clip=true, width=6.76cm]{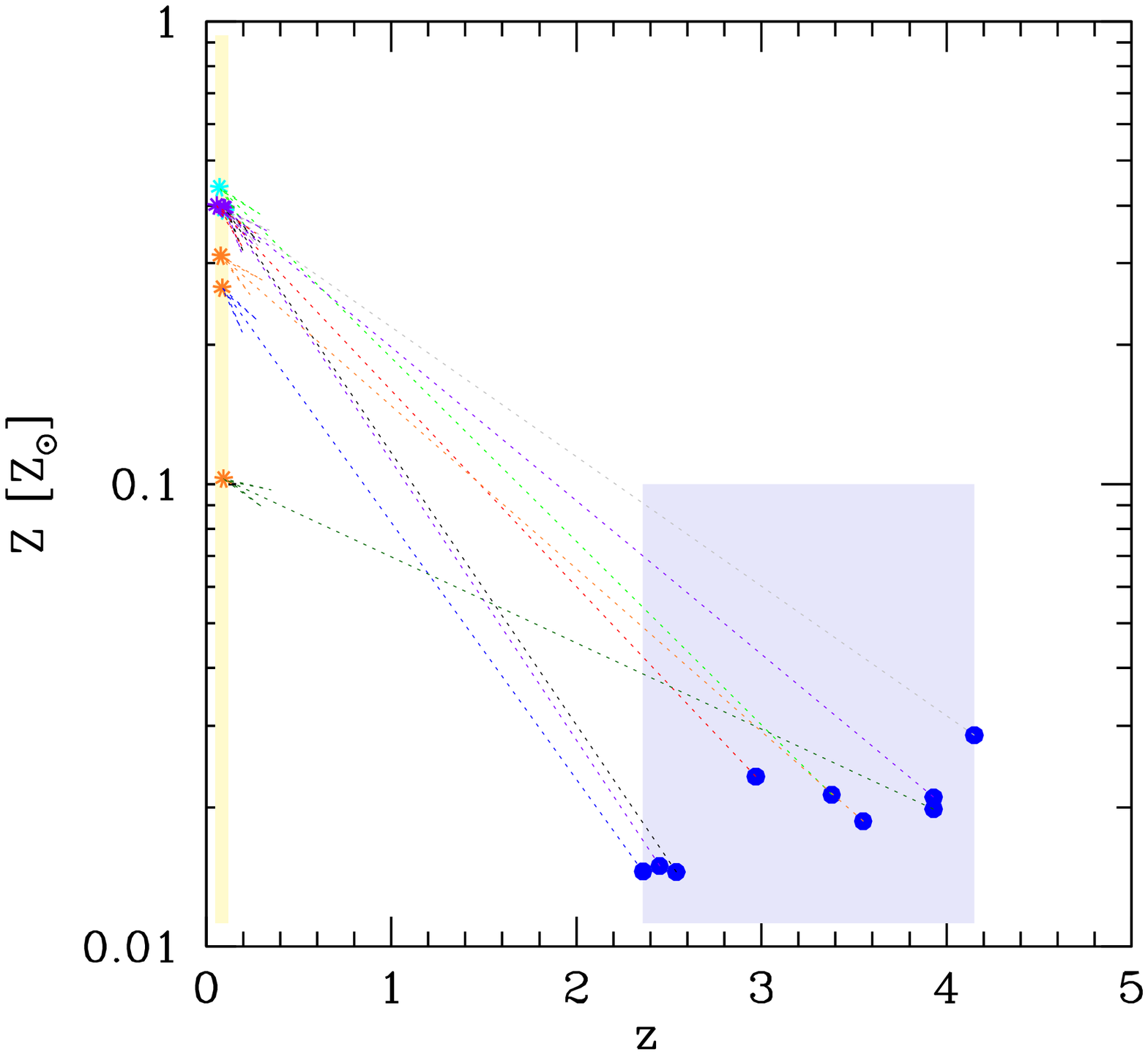}
\vspace{-1.00cm}
\caption{Stellar mass (left), star formation rate (middle), and metallicity (right) of galaxies hosting GW150914-like events. Asterisks show their properties at redshift of BHB coalescence while blue dots indicate the corresponding properties of the galaxy ancestor at the binary system formation. Cyan and light-yellow 
               shadow areas represent the formation ($2.36\leq z_f \leq4.15$) and coalescence ($0.057\leq z_c \leq 0.103$) 
               redshift ranges. Different galaxies are classified as E-, F- and MW systems depending on their 
               dynamical state. Formation halos are connected to the coalescence ones by arrows.}
\label{fig:GW_FORM_HOSTS} 
\end{figure*}
\end{flushleft}
As the MW dominates the global gravitational potential of the simulated volume, the dynamical evolution of the chosen   halos (i.e. both position and velocity of center of mass) is important to determine their fate. The presence of a central attractor in a small cosmic volume does not imply that all small galaxies will be embedded into the central Milky Way by $z_c$ or become close satellites. Our 4 cMpc box is in fact gravitationally constrained by a larger (and less mass resolved) volume of 8~cMpc and many small objects found at its borders are not affected by the global infalling motion. Table 1 shows that only halos residing at a distance of $d_{\texttt{MW}} < 120$~cKpc  at $z_f$ merge into the MW progenitor halo by their $z_c$. MW-systems for example, merge into the MW before $z_c$ and the GW150914-like signal will appear to originate from a MW progenitor galaxy. F-systems, even if gravitationally fragile to tidal interactions, can survive by their $z_c$ as individual galaxies, while only F1 and F3  will remain  at $z=0$ as satellite dwarf galaxies orbiting the MW within $d_{\texttt{MW}} \sim 12$~ckpc. Systems escaping the MW gravitational potential (e.g. E1, E2) are found in our sample at a distance  $d_{\texttt{MW}} > 2.50$~cMpc from the MW at $z_f$. They are mainly affected by the dynamics of the largest  8~cMpc scale and can evolve in isolation, increasing their DM mass by about two orders of magnitude between $z_f$  and $z_c$. In this case the GW150914-like signal will will be hosted by isolated galaxies. 

In summary, among all the potential GW150914 host galaxies (asterisk-shaped data in Figure 1), 44\% merge into the MW halo progenitors by $z_c$, while the remaining ones survive as MW satellites or isolated galaxies. F2, F3 and E-systems are mainly hosted in intermediate-mass halos (i.e. small Lyman-$\alpha$-cooling halos) with DM masses $10 ^{10}$~M$_\odot \lesssim M_{\rm h} \lesssim 2\times10^{11}$~M$_\odot$. F1 is particularly interesting because it evolves under strong tidal interactions: during its infall it experiences mass stripping, its DM halo mass decreases, and $T_{vir} \sim 1.5\times 10^4$~K by $z_c$; F1 becomes then a non star-forming galaxy located at $\sim$700~ckpc from the MW.   
\begin{table*}
\begin{centering}
\begin{threeparttable}
\begin{tabular}{|c|c|c|c|c|cc|c|c}
\hline
{\bf(gID)} Obs.& $ z_c \rightarrow z_s ; z_{obs}$  & Z [12+log(O/H)] &   log(M$_\star$)~[M$_\odot$] & log(SFR)~[M$_\odot$/yr] \tabularnewline
\hline
\hline
{\bf (F1)} UGC4483 \tnote{b} & {$\bm{0.095}$} $\rightarrow \bm{0.0}$ ; $0.0005$   & $\bm{7.53} \rightarrow \bm{7.68}$; $7.46\pm0.02$ \tnote{d}   & 
$\bm{6.86} \rightarrow \bm{6.87}; 6.89\pm0.22$ & {\bf(-)}$ \rightarrow ${\bf(-)}$; -2.21\pm0.18$\tabularnewline
\hline
{\bf(E1)} PGC1446233 \tnote{a} & $\bm{0.072} \rightarrow  \bm{0.02} ; 0.023$    & $\bm{7.48}\rightarrow \bm{8.39}; 8.38$ \tnote{c}  & $\bm{9.15} \rightarrow \bm{9.19}; 9.11\pm0.09$ & $ \bm{-0.98} \rightarrow \bm{-1.05}; -0.94\pm0.28$\tabularnewline
\hline
\end{tabular}
		\begin{tablenotes}
			\item[a] \cite{2017A&A...604A..53C}.
			\item[b] \cite{2015A&A...582A.121R}.
			\item[c] Calibration from \cite{2013A&A...559A.114M}.
     		\item[d] \cite{2013A&A...557A..95R}.
		\end{tablenotes}
\end{threeparttable}
\par\end{centering}
\caption{\label{tab:barGalaxies} Observed analogues of the simulated galaxies where GW150914 are predicted to occur. Obs. is the name of the observed galaxy and gID is the unique evolutionary channel identifier. $z_{obs}$ is the redshift of the observed galaxy, $z_s$ is the redshift closer to $z_{obs}$, while $z_c$ is the redshift of coalescence of GW150914-like event. M$_{*}$ is the stellar mass of the simulated/observed galaxy at $z_c$ and $z_s$, while SFR indicates their star formation rate; note that when the galaxy does not form stars the value of log(SFR) is indicated as (-). For an easier reading simulated values are in bold.}
\end{table*}

We summarize the baryonic evolution of the galaxies living in these halos in Figure 2 by showing their stellar mass ($M_{\star}$, left panel), star formation rate (SFR, middle panel) and metallicity ($Z$, right panel), as function of $z$. For an easier comparison, this figure follows the symbol convention of Figure 1, while Table 2 summarizes the galaxy properties.

All the galaxies (but the most massive MW2) have a similar initial stellar mass $M_{\star} \sim 2 \times 10^{6} $~M$_{\odot}$ while the stellar content of their descendants shows a large scatter. This means that their individual  evolution can significantly shape their properties. For example, all the galaxies evolving in relative isolation (i.e. E-systems and F3) can easily grow (reaching $M_{\star} \sim 10^9$~M$_{\odot}$ at $z_c$) because they can avoid strong dynamical interactions. Under these conditions they can accumulate stellar mass and develop a mature ISM as indicated by the increase in their metallicity from  Z$\sim10^{-2}$~Z$_{\odot}$, to $Z \gtrsim 0.4$~$Z_{\odot}$. Also note that their final metallicity is comparable to the one of the MW progenitor, i.e. to fastest growing systems in the simulation box.

F2 and F1, on the other hand, experience a more peculiar evolution. F2 shows an increment of about two orders of magnitude in its stellar mass (note that the corresponding increase in DM mass is one order of magnitude) while having a final SFR comparable to the value at $z_f$. This indicates that between $z_f$ and $z_c$, F2 experienced a phase of efficient star formation, which was successively quenched as the halo entered the potential of the MW and stopped accreting gas to fuel star formation. Note that the corresponding increase in metallicity is of about one order of magnitude, i.e. from $Z\sim0.01$~Z$_{\odot}$ solar to $Z\sim0.3$~Z$_{\odot}$.
 
The history of F1 reveals an even more complex interplay of feedback processes: it is continuously stripped in DM mass during its infalling path and reaches the potential of a mini-halo at $z_c$. This makes its capability to form stars extremely bursty and limits its total stellar mass to $M_{\star} \sim 8\times 10^6$~$M_{\odot}$. At $z_c$ this galaxy is not forming stars (SFR = 0, note that its row in Figure 2 is absent) and it has the lowest metallicity of our sample (about a factor of three lower).

The assembly of the MW progenitor galaxies has been described in detail in GR17 and it is not discussed in this section; we then refer the interested reader to this work.

\subsection{Dwarf galaxy analogues}

We searched for observable counterparts of the dwarfs discussed in the previous section within the observational samples described in Section \ref{sec:sample}.  
To this aim, a two steps searching  algorithm was adopted: first we built an initial sub-sample by selecting galaxies that match (within the observational uncertainties) at least two quantities in the $M_\star$-SFR (the so-called main-sequence), $M_\star$-$Z$ or SFR-$Z$ planes. Second, we picked out only the observed analogues having the third quantity differing by less than a factor of two. The resulting galaxies are then identified as  plausible observable counterparts of dwarfs hosting GW150914-like events. Note that when candidate galaxies are observed at $z_{obs} $ slightly different from $z_c$, we compare their properties with the descendant hosts found at simulation snapshot having a redshift $z_s$ closer to $z_{obs}$.\newline
Table \ref{tab:barGalaxies} summarizes properties of both simulated and observed galaxies matching the above requirements. For an easier reading simulated values are in bold. We find for example that descendants of E1 and F1 have properties consistent with the two systems PGC1446233 (from the ALLSMOG survey) and UGC4483 (from the DGS sample).
In particular, UGC4483 is a nearby  Blue Compact Dwarf (BCD) galaxy (see \citealp{2012A&A...544A.145L}) with very low stellar mass and metallicity ($ M_\star \sim 7.7\times 10^7 M_\odot$, $Z\sim 0.06 Z_\odot$). It is interesting to note that UGC4483 shows extreme properties with respect to other galaxies of the same catalogue. Its total dust mass, for example, is difficult to reproduce: the observed value is $M_{\rm{dust}}\sim 280 ~ M_{\odot}$ \citep{2015A&A...582A.121R}, almost one order of magnitude smaller than the dust mass expected (even assuming only the contribution from stellar sources, \citealt{2018MNRAS.473.4538G}) for its simulated analogue F1. Consistently with the discussion in the previous section, UGC4483 can be interpreted as a galaxy evolving under strong feedback, and showing a poorly evolved interstellar medium, with an unstable ISM molecular gas phase. Note on the other hand, that the F1/UGC4483 dust mass excess is consistent with the hypothesis of efficient dust destruction by SN shocks in our models, as detailed in \cite{2018MNRAS.473.4538G}.

Finally note that the above comparison with observational data has the primary goal to show that our model predicts extremely peculiar, low metallicty objects which are not model artifacts but have observed counterparts. Also note that the simulated sample refers to a small, well resolved  cosmological volume and it is not  suitable to compute coalescence rates. This would require a larger cosmological volume with the same resolution to provide a statistical sample of both massive and dwarfs galaxies but to our knowledge it is beyond the current computational capabilities; we then defer this point to a future investigation.

\begin{flushleft}
\begin{figure*}
\hspace{-1.0cm}
\includegraphics [clip=true, width=6.76cm]{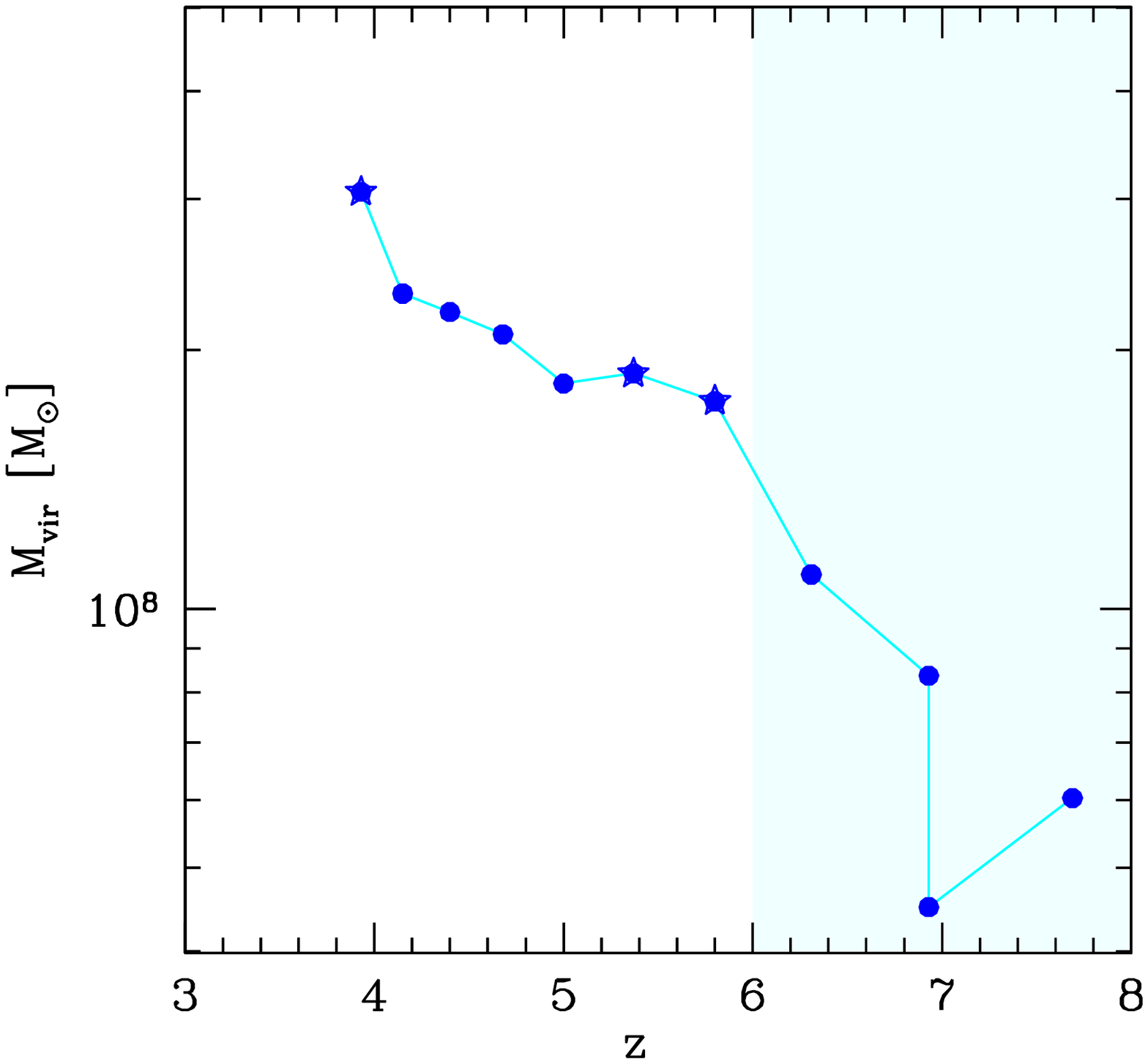}
\hspace{-1.05cm}
\includegraphics [clip=true, width=6.76cm]{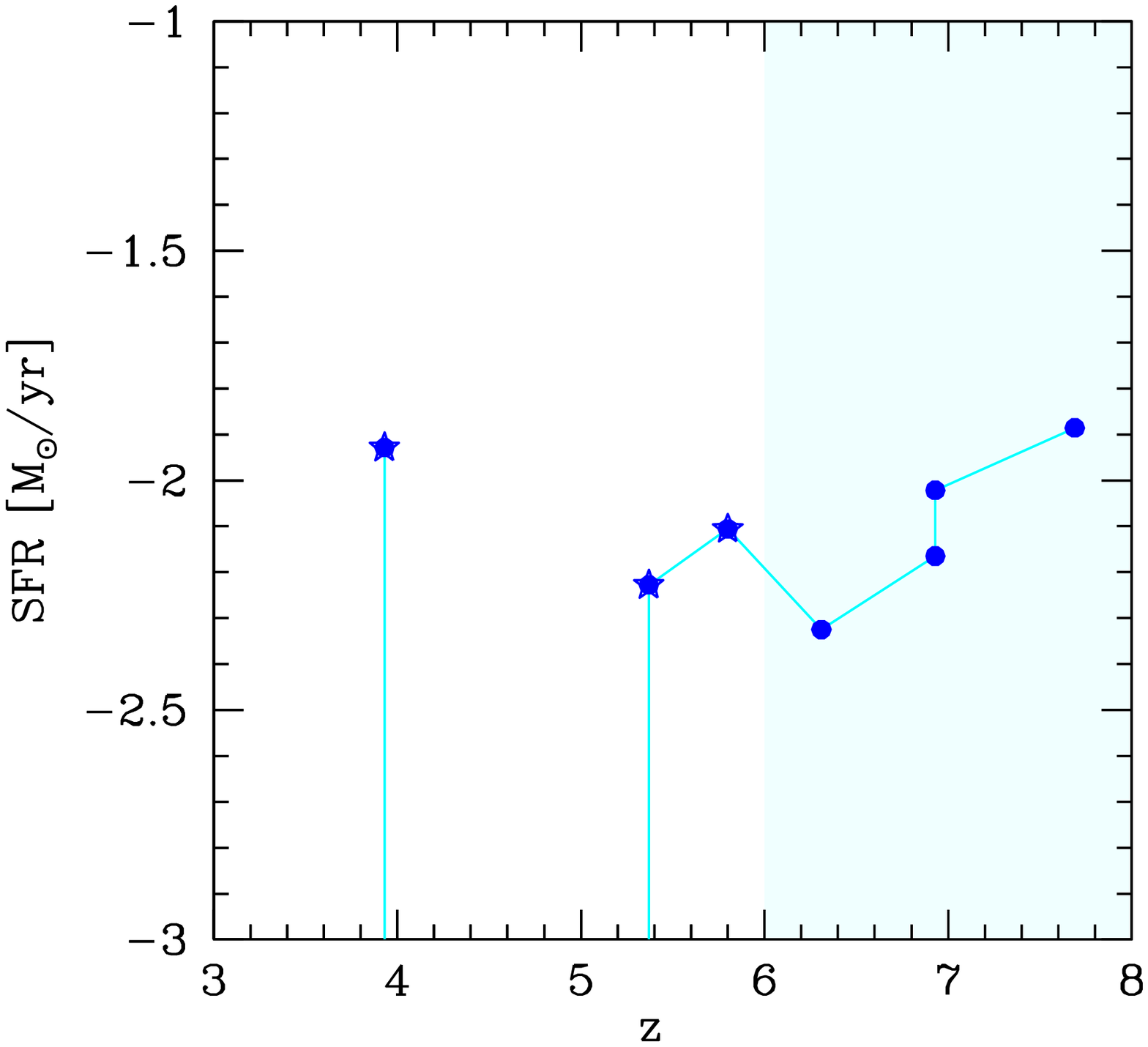}
\hspace{-0.99cm}
\includegraphics [clip=true, width=6.76cm]{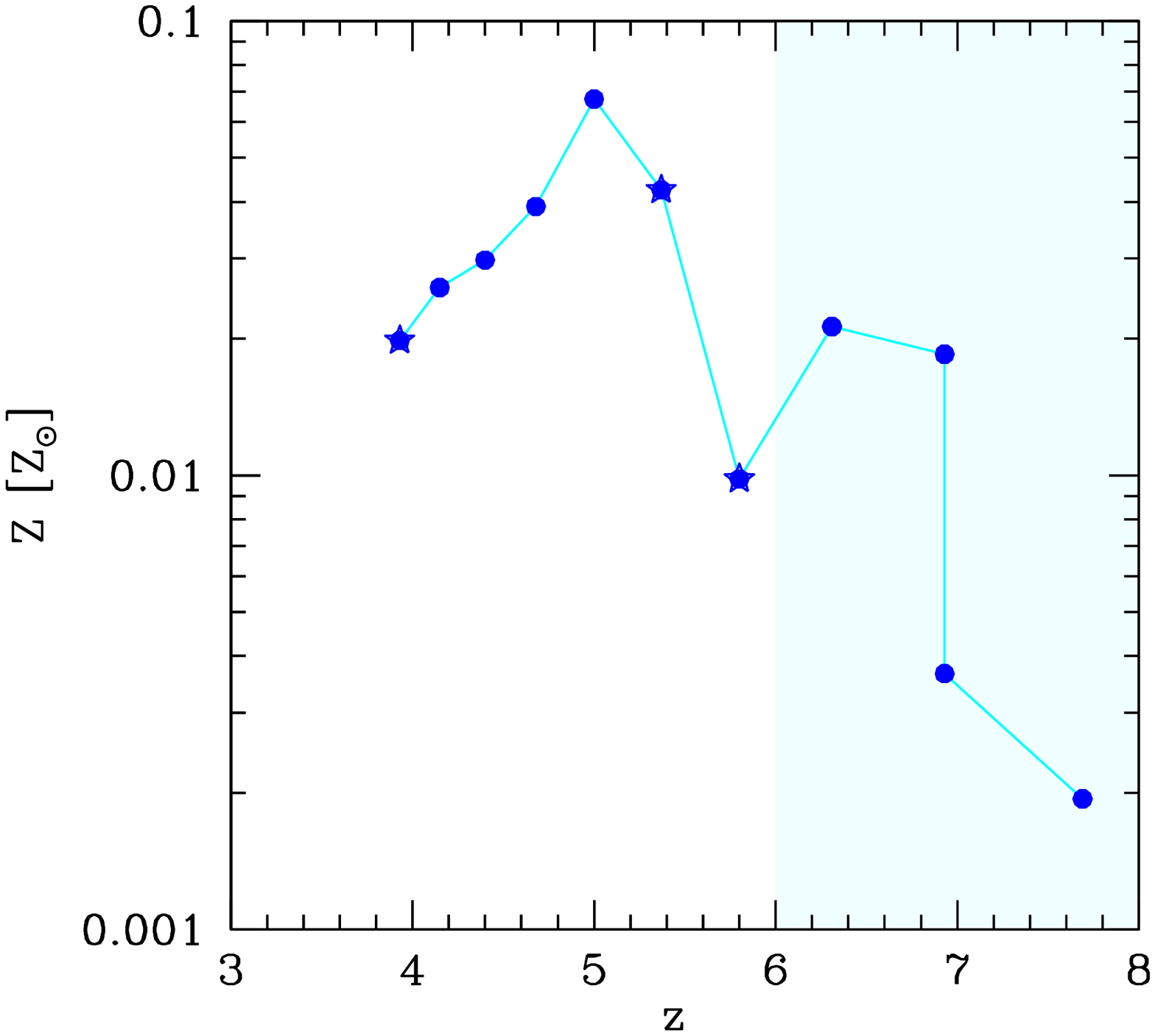}
\vspace{-1.1cm}
\caption{Evolution of the channel F1 from its formation, down to $z_f=3.93$, when the candidate binary systems form. {\bf Left panel}: Redshift evolution of DM halo virial mass $M_{vir}$ . {\bf Middle panel}: Redshift evolution of the galaxy  SFR~[M$_{\odot}$/yr]. {\bf Right panel}: Redshift evolution of the gas metallicity $Z \leq 0.5 Z_{\odot}$.}
\label{fig:GW_FORM_HOSTS_F1} 
\end{figure*}
\end{flushleft}

\subsection{Early evolution of formation hosts}

This section explores the most interesting evolutionary histories (hereafter also referred as evolutionary channels)  of E- and F-galaxies hosting GW150914-like events. The history of MW-like progenitors found  in the same simulation is discussed in detail in GR17 and we refer the interested reader to this publication. As discussed in the previous section the redshift $z_f$ at which these galaxies form the candidate binaries is spread in a wide range ($2.36 \leq z_f \leq 4.15$) and it is theoretically important to understand the diversity of histories leading to these peculiar environments at $z_f$. 

In Figures \ref{fig:GW_FORM_HOSTS_F1}, \ref{fig:GW_FORM_HOSTS_F2} and \ref{fig:GW_FORM_HOSTS_E1} we explore the three most informative channels found in our simulation by showing the redshift evolution of the DM halo virial mass (M$_{vir}$~[M$_{\odot}$], left panel), the galaxy star formation rate (SFR~[M$_{\odot}$/yr], middle panel) and the galaxy metallicity ($Z$~[$Z_{\odot}$]}, right panel). In all the panels, the epoch of reionization (i.e. $z>6$ in our model) is indicated by a cyan shadow area. Blue dots indicate when the DM halo is a mini-halo (i.e. cannot form stars after reionization) while the star-shaped points indicate when the galaxy is hosted in a Lyman-$\alpha$-cooling halo (i.e. it can form stars at any redshift because it is unaffected by radiative feedback). When the candidate galaxy has more than one progenitor, they are shown by different line colours and their separate evolution is referred as evolutionary channel (CH). When two channels converge into a single one, the halos/galaxies experience a merger event. 

Figure \ref{fig:GW_FORM_HOSTS_F1} shows the early evolution of F1, from the redshift of the DM halo collapse ($z_{hc} \sim 7.7$), down to $z_f\sim 3.93$. The halo forms at a comoving distance from the MW $d_{\texttt{MW}}\sim 2.7$ cMpc, with an initial mass of $M_{vir}\sim 6 \times 10^7$~[M$_{\odot}$] and immediately experiences a strong tidal interaction which decreases its value by  $\sim 15$\%. After this initial event, it continues to smoothly grow in isolation without exchanging significant mass with nearby objects, while falling toward the MW. Note that in the above redshift range its distance from the MW changes only by 300~ckpc, indicating a low infall rate. Its galaxy is allowed to form stars at $z>6$ because its DM halo slowly virializes by accreting mass from the surrounding IGM and remains a mini-halo for large part of the early evolution\footnote{Note that this depends on the peculiar scheme of radiative feedback adopted in this simulation. When a fully radiative transfer scheme is adopted this prediction can change depending on the topology of the in-homogeneous reionization. This scenario will be explored in a future work.}: its initial SFR is very low (SFR$\sim 10^{-2}$~[M$_{\odot}$/yr]) and decreases down to SFR$\sim 4 \times 10^{-3}$~[M$_{\odot}$/yr] because it rapidly exhausts the available gas acquired from the IGM. Down to $z\sim 5.5$ the SFR is strongly affected by radiative feedback because the halo has a virial temperature only episodically higher than $T_{vir} \sim 2\times 10^4$~K and its star formation is definitely suppressed at $z < 5.5$ until the binary candidates form at $z_f\sim 3.93$ when the halo has acquired sufficient mass to become dynamically stable (i.e. a Lyman-$\alpha$-cooling halo) and can form stars again. The redshift evolution of the gas metallicity reflects the  complex interplay between dynamical growth and radiative feedback described above: before reionization the galaxy experiences an irregular increase of its gas metallicity due to the balance between decreasing SFR and increasing mass accretion, which slowly accumulates atomic metals in the gas phase until a value of $Z\sim 7\times10^{-2}$~[Z$_{\odot}$] is reached at $z\sim 5.0$. The successive long evolution is spent increasing its gas content without producing new metal mass; its metallicity decreases then down to $Z\sim 2\times 10^{-2}$~Z$_{\odot}$, allowing the GW150914 candidate to form at $z_f=3.93$.
\begin{flushleft}
\begin{figure*}
\hspace{-1.0cm}
\includegraphics [clip=true, width=6.76cm]{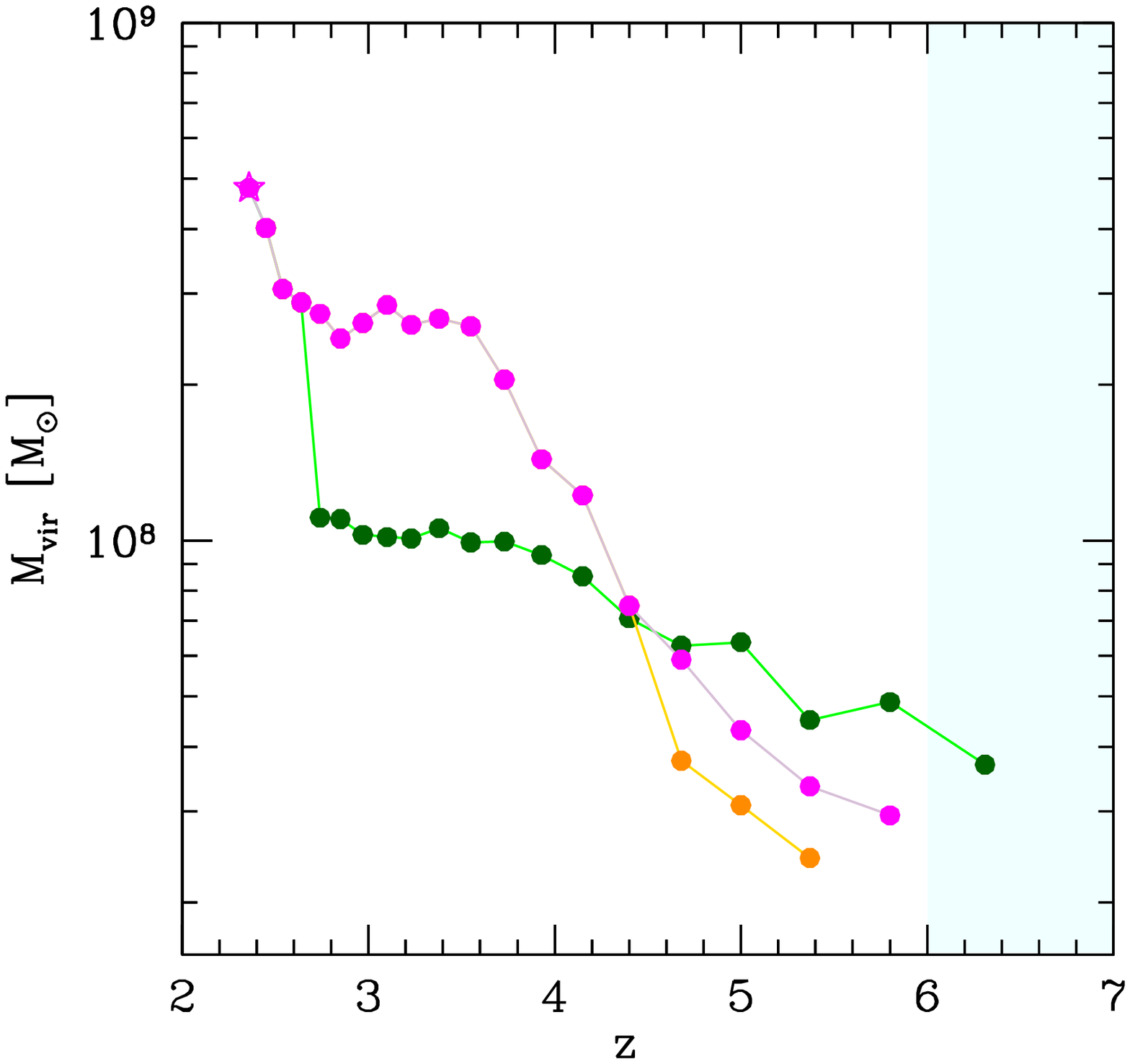}
\hspace{-0.99cm}
\includegraphics [clip=true, width=6.76cm]{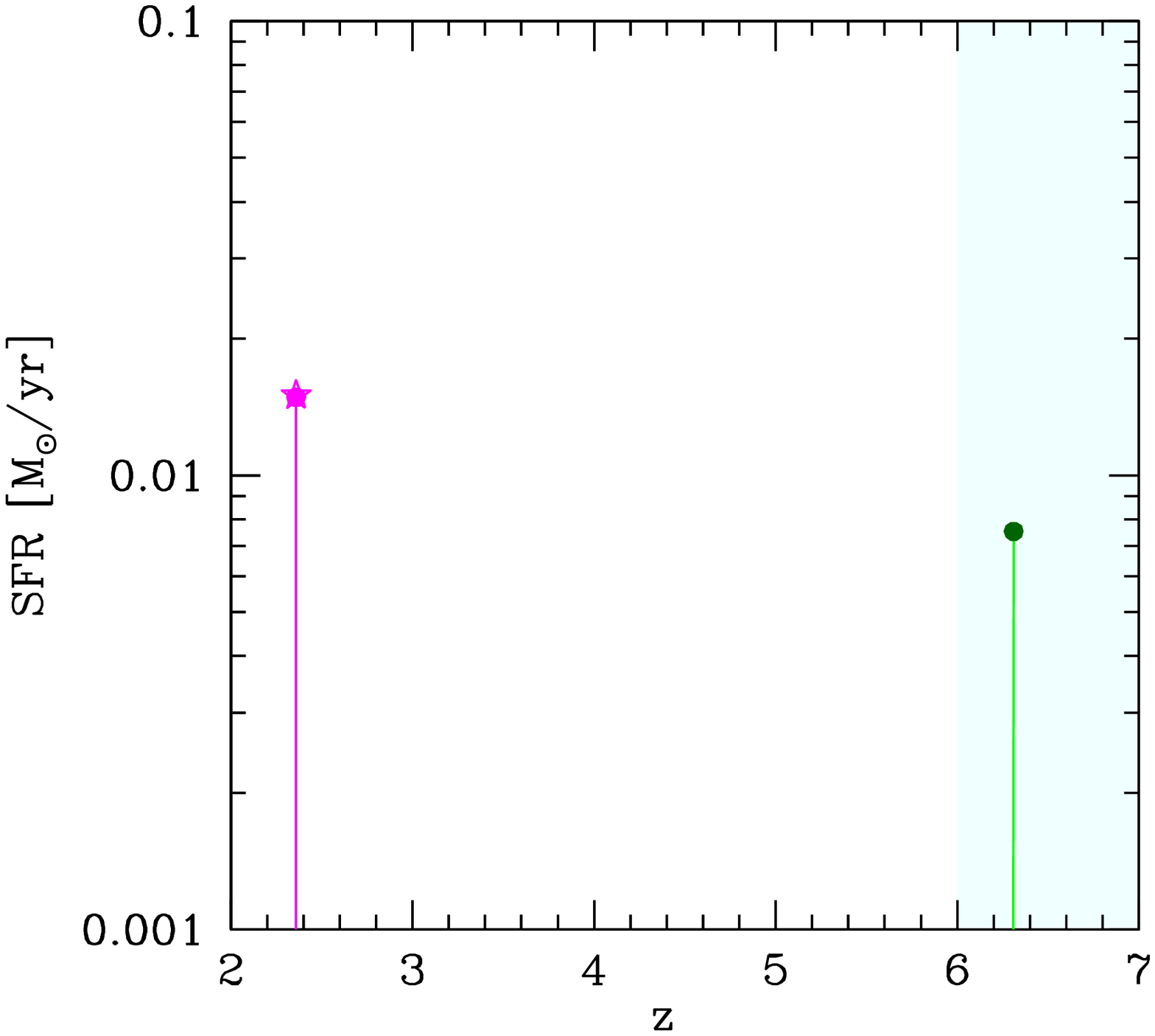}
\hspace{-1.10cm}
\includegraphics [clip=true, width=6.76cm]{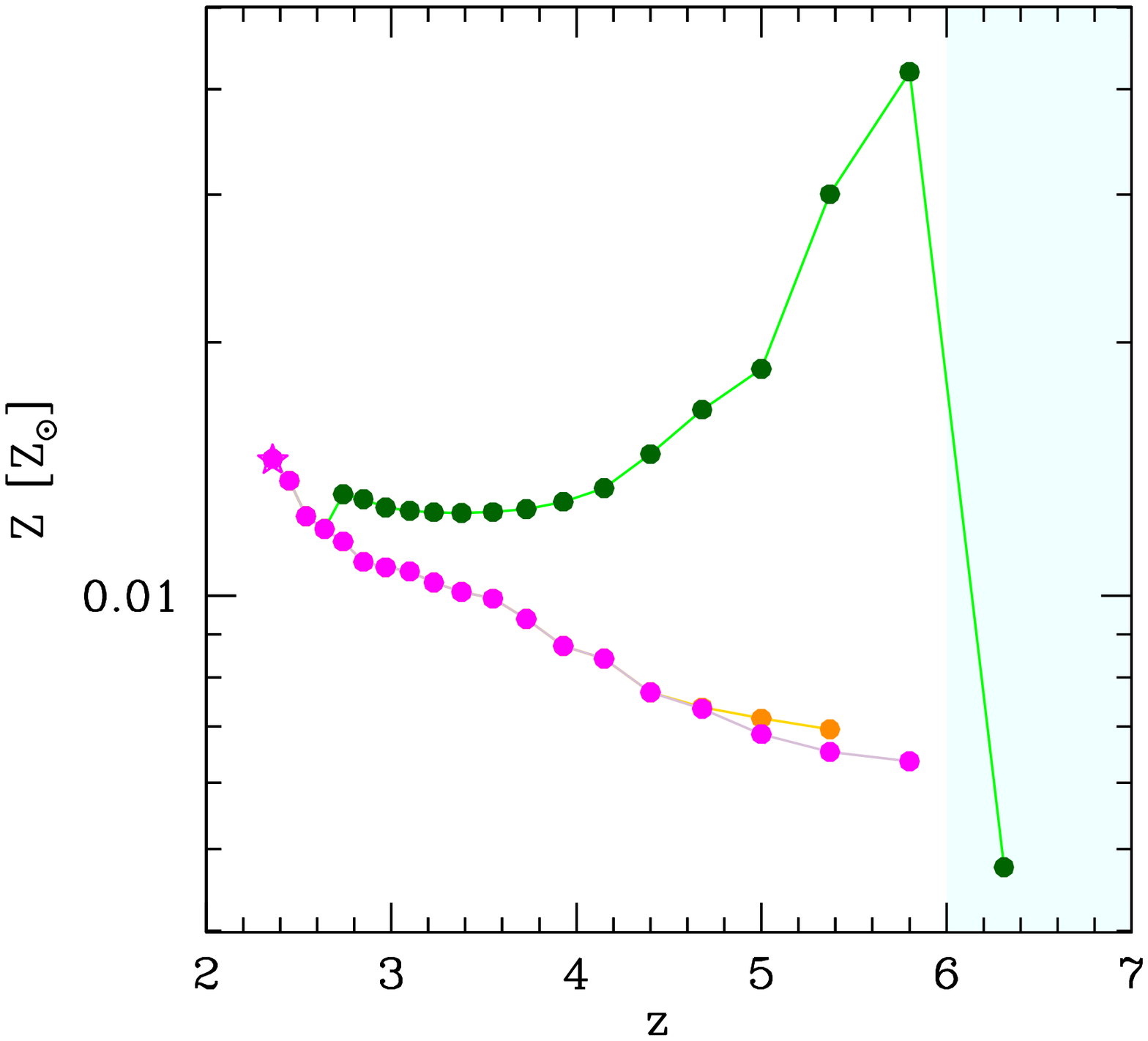}
\vspace{-1.1cm}
\caption{Evolution of channels ending in the galaxy/DM halo F2 at $z_f=2.36$, when the BHB forms. {\bf Left panel}: Redshift evolution of DM halo virial mass $M_{vir}$~[M$_{\odot}$] . {\bf Middle panel}: Redshift evolution of the halo galaxy SFR~[M$_{\odot}$/yr].  {\bf Right panel}: Redshift evolution of the halo metallicity $Z$~[Z$_{\odot}$]. Different colours mark the evolution of the three channels ending in the halo of the BHB binary.}
\label{fig:GW_FORM_HOSTS_F2} 
\end{figure*}
\end{flushleft}
\begin{flushleft}
\begin{figure*}
\hspace{-1.0cm}
\includegraphics [clip=true, width=6.74cm]{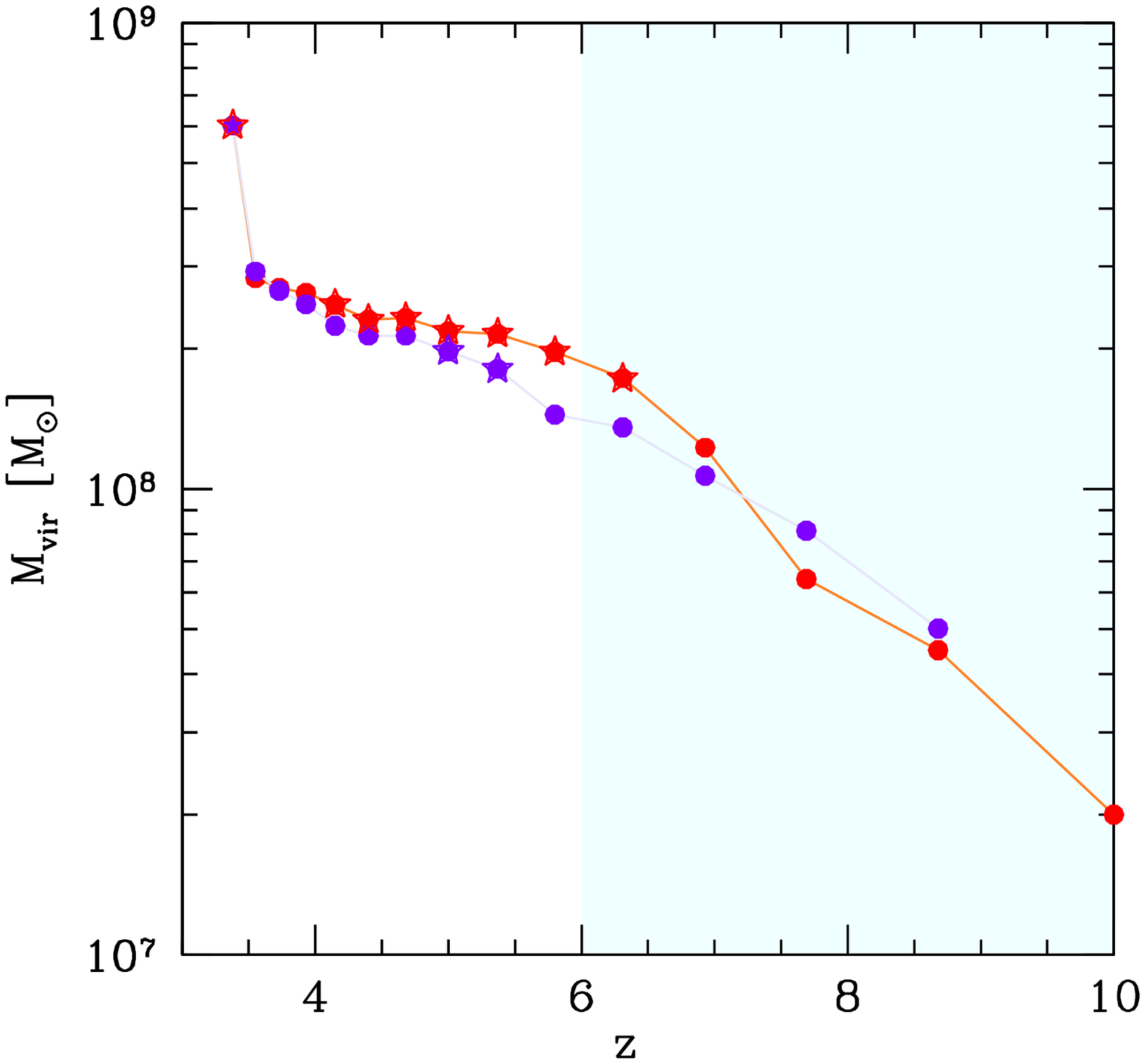}
\hspace{-0.99cm}
\includegraphics [clip=true, width=6.74cm]{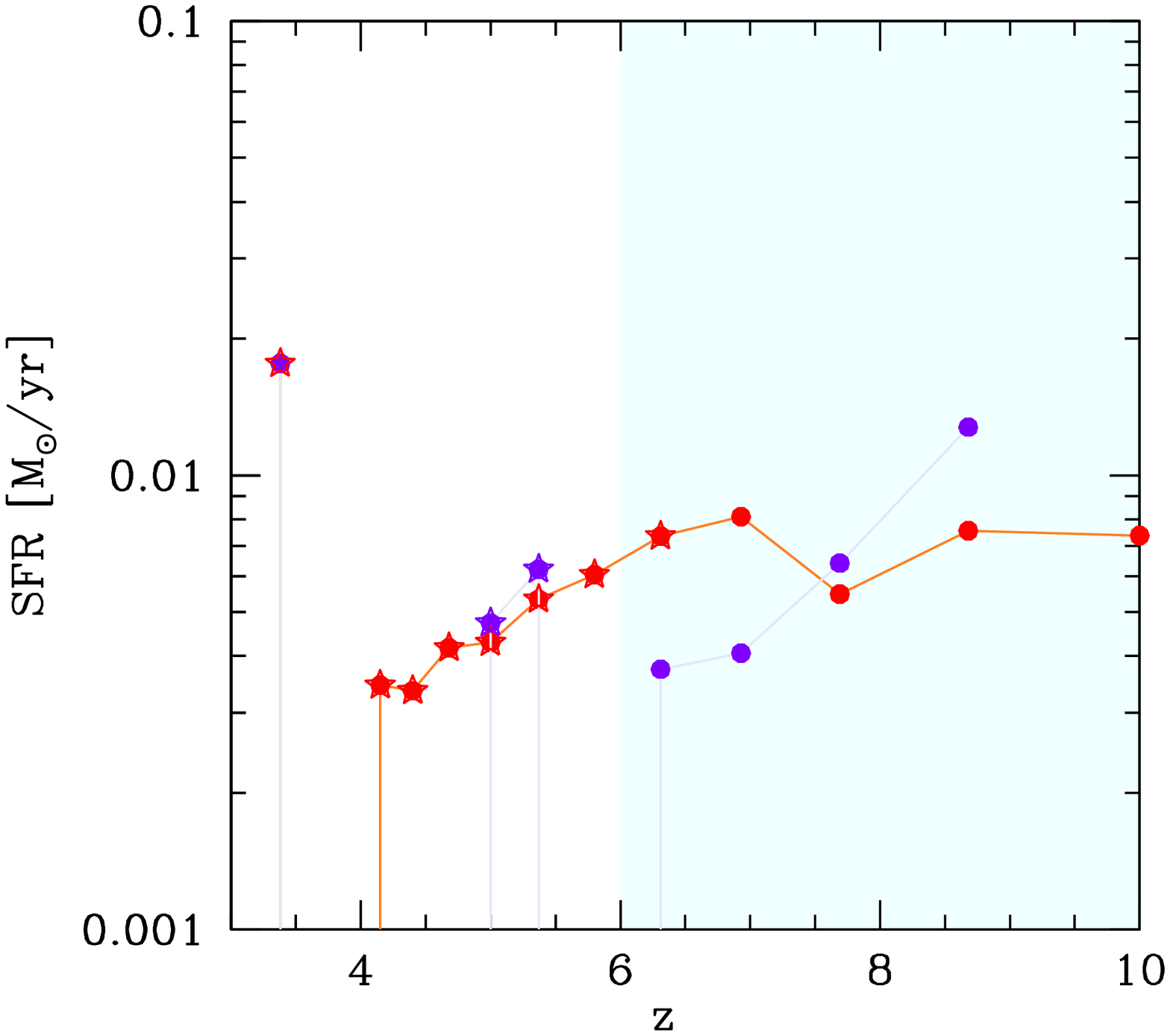}
\hspace{-0.99cm}
\includegraphics [clip=true, width=6.74cm]{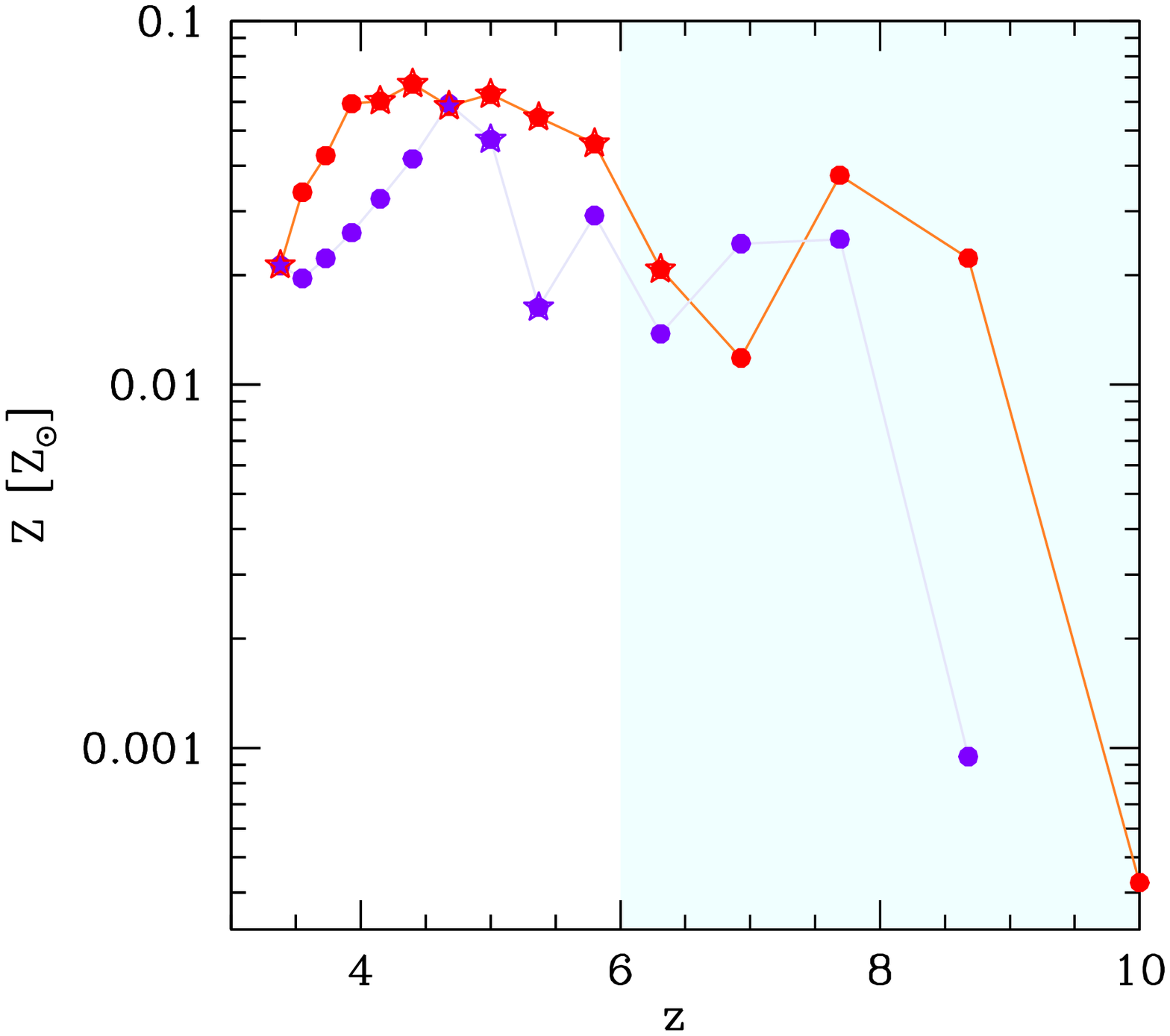}
\vspace{-1.1cm}
\caption{Evolution of channels ending in the galaxy/DM halo E1 at $z_f=3.38$, when the BHB forms. {\bf Left panel}: Redshift evolution of DM halo virial mass $M_{vir}$~[M$_{\odot}$] . {\bf Middle panel}: Redshift evolution of the halo galaxy SFR~[M$_{\odot}$/yr]. {\bf Right panel}: Redshift evolution of the halo metallicity $Z$~[Z$_{\odot}$]. Different colours mark the evolution of the two channels ending in the halo of the BHB binary.}
\label{fig:GW_FORM_HOSTS_E1} 
\end{figure*}
\end{flushleft}

The early evolution of F2 is shown in Figure \ref{fig:GW_FORM_HOSTS_F2} and results from the assembly of three mini-halos collapsed at different redshifts ($z_{hc}\sim 6.4$ (green), $5.8$ (violet), $5.4$ (orange)) but at a common initial comoving distance of $d_{\texttt{MW}}\sim 2.4$~cMpc from the MW progenitor. The redshift of formation has a strong influence on their successive evolution: two channels of F2 start at $z_{hc} < 6$ and can not form stars because they are suppressed by radiative feedback before their merging event. A single episode of star formation (SFR$\sim 8\times10^{-3}$~M$_{\odot}$/yr) is allowed instead in the remaining channel only because it collapsed before reionization. As a result, the metallicity of the final halo at $z_f=2.36$ is very low (Z$ \sim 10^{-2}$~[Z$_{\odot}$]) and it is largely regulated by the accretion of metal-enriched gas.

E1 shows an alternative assembly history (see Figure \ref{fig:GW_FORM_HOSTS_E1}): it is created by two halos evolving separately from $z_{hc}\sim10$ (red line, CH1) and $z_{hc} \sim 8.5 $ (blue line, CH2) and merging at $z_f=3.38$. Both halos are created with masses $M_{vir} \sim 2\times 10^7$~$\rm M_{\odot}$ and $M_{vir} \sim 5 \times 10^7$~$\rm M_{\odot}$ respectively, and grow with a similar trend but dynamically separated: CH1 does not follow the global infall and stays along its evolution at a fixed distance $d_{\texttt{MW}}\sim 2.5$~cMpc from the MW. CH2 is attracted by CH1 when following the global infall. CH1 (the first virializing halo) becomes a Lyman-$\alpha$-cooling type before the end of reionization and then continues to form stars during its successive evolution, while CH2, being more fragile, forms stars only before $z\sim 6$ (i.e. before reionization completes). It successively experiences only sporadic episodes of star formation before merging into CH1. After their galactic merger, the first episode of star formation occurs at a rate  SFR$\sim 2 \times 10^{-2}$~[M$_{\odot}$/yr] and also creates the candidate GW150914 system in  galactic environment having a gas metallicity Z$ \sim 2 \times 10^{-2}$~Z$_{\odot}$. Note that the dynamical interactions occurring before the merging event have a negative impact on star formation by changing their virial temperature, until the merger occurs. The result of this combined feedback is also reflected in the evolution of their metallicity: it naturally increases when the systems are star forming and successively decreases until the final merger.

\section{Discussion and Conclusions}
\label{sec:Discussion} 
The identification of galaxies hosting gravitational wave events is a scientific problem of increasing interest \citep{2013ApJ...767..124N,2014ApJ...795...43F,2016arXiv161201471C} and it will be at the core of future investigations in gravitational astronomy for the following, primary reasons: (i) the redshift of the GW host galaxy provides an independent estimate of the emission epoch of the GW event and will allow to use gravitational wave sources as standard cosmological sirens \citep{1986Natur.323..310S, 2012PhRvD..86d3011D}; 
(ii) observations of the host galaxy provide constraints on the formation mechanism of the coalescing binary 
(\citealt{2014ApJ...795...43F,2016ApJ...818L..22A} and references therein).

In the case of GW150914, different collaborations developed strategies to search for electromagnetic transients but so far, no counterparts consistent with the LIGO detection have been identified  (see \citealt{2017arXiv171005915B,2016MNRAS.462.3528C,2016ApJ...828L..16D,2016ApJ...823L...2A}, and references therein). The Fermi Collaboration does not detect any transient events associated with GW150914 \citep{2018ApJ...853L...9C}. Observations with the INTEGRAL and MAXI satellites have not detected any counterpart of the LIGO event \citep{2016ApJ...820L..36S}. It should be noted that even if mergers of BHs from stellar binaries are usually not supposed to have any electromagnetic counterpart, under specific condition some circumstellar material may be left after BH merger and could generate a signal \citep{2016ApJ...819L..21L}.

In GW150914-like scenarios, in which the electromagnetic counterpart is supposed to be absent or weak, the determination of the redshift of the galaxy hosting coalescence and a theoretical estimate of the epoch of the binary formation become then of primary importance \citep{2013ApJ...779...72D}. As we have shown in the present work, these questions require a theoretical framework capable to follow the binary systems consistently with the redshift evolution of their environments.
An additional complication at this stage of our model, consists in the fact that we are not able to remove the degeneracy 
in our nine potential hosts and we can not identify a more {\itshape probable candidate}. This is because {\itshape all} the hosted binary systems coalesce in the redshift range of LIGO \citep{2016PhRvL.116f1102A}.
In principle this degeneracy could be removed by additional constraints on the physical properties of the BHB sources, 
for example by including the spin of the two black holes in our models. At present the GW observation of a single coalescing BHB can not strongly constrain individual BH spins \citep{2016ApJ...818L..22A}; this leaves a significant degree of uncertainty in spin estimates. However taking into account spins could be very useful to disentangle spin distribution scenarios \citep{2010CQGra..27k4007M,2017arXiv170907896F}. We defer this important point to a future investigation adopting an improved 
version of \texttt{SeBa}  capable to follow the spin evolution of the BHBs along the cosmological evolution of the \texttt{GAMESH} galaxies (Marassi \& Graziani in prep.).

In this work we found that 
GW150915, one of the most massive binary event detected so far (see Table III in \citealt{2018arXiv181112907T}), can be hosted in a small number 
of representative galaxies of our local cosmic volume. Thanks to this exiguous sample, we
have been able to follow the evolutionary histories of single dwarfs selected as potential GW150914 hosts. Among them we have recognized three classes of halos hosting GW150914-like events: (i) MW-systems belonging to 
the main MW progenitor; (ii) F-systems belonging to MW satellites; (iii) E-systems halos hosting 
dwarf galaxies that can dynamically escape the global infall onto the central Milky Way.  
 
We have shown that:

\begin{itemize}

\item the difference in individual histories and the complexity of environmental effects imply that among 13000 available dwarfs, only nine representatives are able to host GW150914-like event.
 
\item among them, only $44\%$ of the galaxy candidates merge into the MW halo progenitors before the redshift of coalescence of their BHBs. The remaining galaxies are found as isolated or satellites dwarfs from which the signal could originate (but see \citealt{2017ApJ...850L...4C} for an alternative scenario); their evolution is strongly affected by both dynamical history of DM halos and environmental feedback;

\item by comparing the physical properties of these candidates with observational catalogues of the DGS survey 
\citep{2013PASP..125..600M} and  in ALLSMOG survey \citep{2017A&A...604A..53C} we found an extremely good agreement with two observed candidates (PGC1446233 and UGC4483). These clearly show signatures of a peculiar history and are highly shaped by feedback effects;

\item we finally investigated different evolutionary histories leading to the right conditions to form GW150914-like binaries in the high-redshift universe. Alternative channels show that an evolution in isolation is not strictly required to meet a low metallicity condition at the redshift of the binary formation. A more stringent condition comes instead from a combination of their accretion history and radiative feedback in the early universe, in which they usually collapse. We then emphasize the unique role of radiative feedback in determining the early, individual star formation and enrichment histories. This point will be addressed in a future work, and in a more appropriated context of in-homogeneous reionization and spatially dependent metal enrichment. 

\end{itemize}

Finally, we highlight that the novel approach adopted in our work can be applied to all the families of compact binaries predicted by different population synthesis codes, as double neutron stars and black hole-neutron stars binaries. Also consider that the very recent announcement of \citet{2018arXiv181112907T} reported the detection of new four massive coalescing black holes in the redshift range $0.20\leq z\leq 0.48$: GW170729 ($\rm M_1 \sim 50.6$~$M_\odot$, $\rm M_2 \sim 34.3$~$M_\odot$), GW170809 ($\rm M_1 \sim 35.2$~$M_\odot$, $\rm M_2 \sim 23.8$~$M_\odot$), GW170818 ($\rm M_1 \sim 35.5$~$M_\odot$, $\rm M_2 \sim 26.7$~$M_\odot$), GW170823 ($\rm M_1 \sim 39.6$~$M_\odot$,$\rm M_2 \sim 29.4$~$M_\odot$). In light of this new data-set, a future study will enlarge our statistic sample (see Appendix B for more details) and consequently extend our analysis to account for a wider statistic in our simulations. 

\section*{Acknowledgments}
The authors would like to thank Silvia Piranomonte and the internal referee of the LIGO-VIRGO collaboration for useful suggestions. 
We also acknowledge PRACE 
\footnote{http://www.prace-ri.eu/} for awarding us access to the CEA HPC facility "CURIE@GENCI"
\footnote{http://www-hpc.cea.fr/en/complexe/tgcc-curie.htm} with the Type B project: High 
Performance release of the GAMESH pipeline.
MM  acknowledges financial support from the MERAC Foundation through grant `The physics of gas and protoplanetary discs in the Galactic centre', from INAF through PRIN-SKA `Opening a new era in pulsars and compact objects science with MeerKat' and from the Austrian National Science Foundation through FWF stand-alone grant P31154-N27 `Unraveling merging neutron stars and black hole - neutron star binaries with population-synthesis simulations'.  This work benefited from support by the International Space Science Institute (ISSI), Bern, Switzerland,  through its International Team programme ref. no. 393
 {\it The Evolution of Rich Stellar Populations \& BH Binaries} (2017-18).
The research leading to these results has received funding from the European 
Research Council under the European Union's Seventh Framework Programme 
(FP/2007-2013) / ERC Grant Agreement n. 306476.

\bibliographystyle{mnras}
\bibliography{GAMESH2GWHalos}{}

\begin{appendix}

\section{The GAMESH model of Galaxy formation}
\label{sec:GAMESH_MODEL} 
\texttt{GAMESH} \citep{2015MNRAS.449.3137G} is a galaxy formation model based on a hybrid pipeline which consistently combines three modules: the outputs of a DM simulation, a semi-analytic model (SAM) of star formation and chemical feedback and a Monte Carlo-based radiative transfer algorithm (see Figure 1 in the original publication). 

\begin{figure*}
\hspace{-1.10cm}
\includegraphics [clip=true, width=8.0cm]{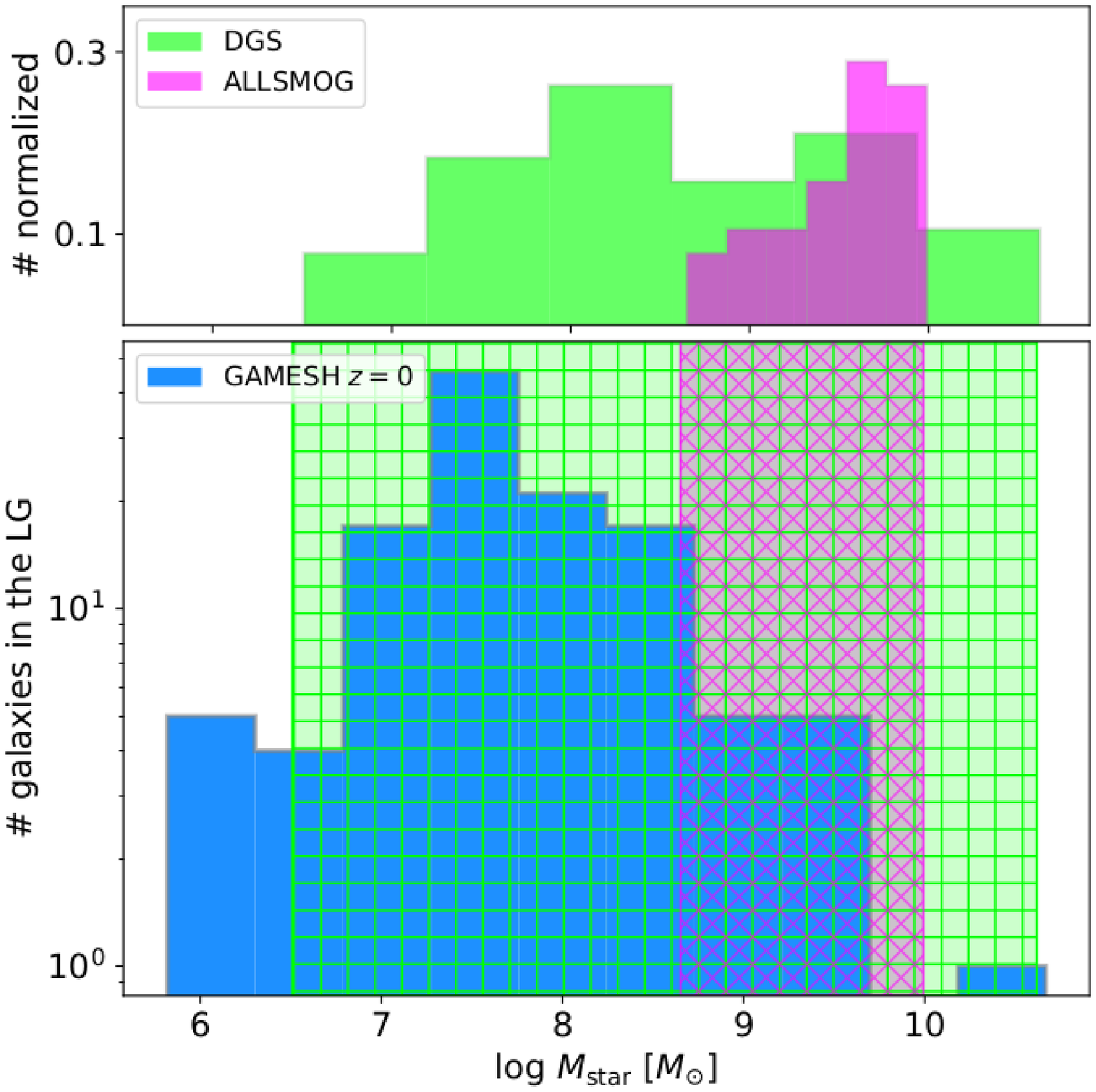}
\includegraphics [clip=true, width=8.0cm]{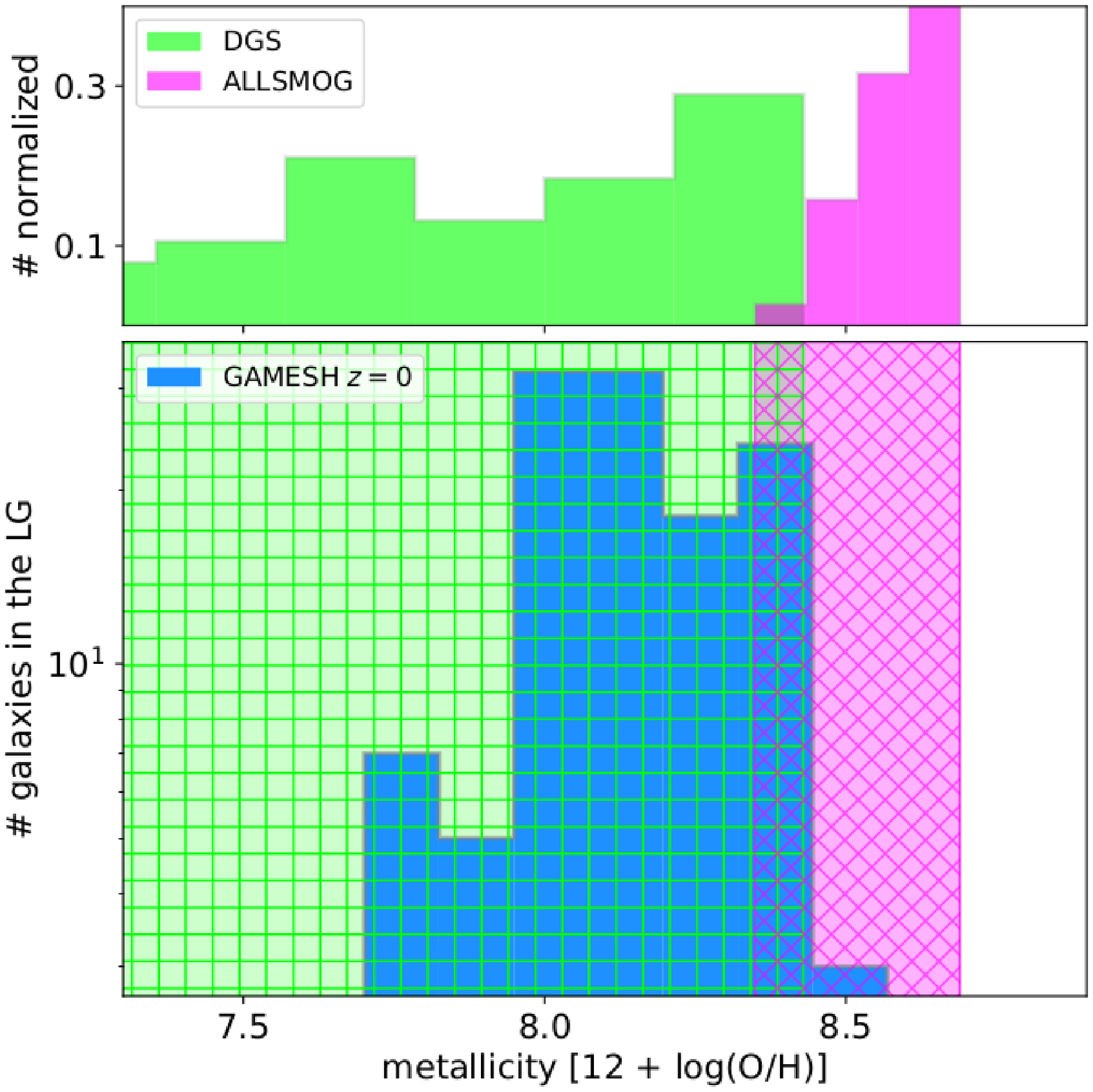}
\hspace{-1.10cm}
\caption{Statistic of galaxies in the LG as function of their log(M$_{\star}$) (left panel) and metallicity (right panel). The blue histograms represent simulated data at $z=0$ while the range of values found in the DGS and ALLSMOG are shown as green and violet areas, respectively. The top panels of the same figure show the normalized histograms of the observed samples with same colour conventions.}
\label{fig:GAMESH_DWARFS} 
\end{figure*}

A typical run of \texttt{GAMESH} is based on a fixed DM simulation which describes the evolution of structures on a series of snapshots at assigned reshift $z_i, i \in {1,..N}$, where $z_N=0$. At each $z_i$ the DM simulation provides the spatial distribution of the matter\footnote{The gas mass at $z_1$ is derived by scaling the DM mass with the universal baryon fraction.} in the cosmic volume and the catalogue of halos, while the SAM simulates the processes of star formation and stellar evolution (i.e. evolution of stellar populations, their mechanical feedback from winds and the chemical enrichment caused by supernova explosions)\footnote{Here we point out that the galaxy baryonic properties (e.g. mass of gas, stars and metals) are derived in each halo without adopting any scaling relation or abundance matching tecnique but by creating new stellar mass from the available gas and by following the feedback from its stellar populations.}. The mass exchange with the intergalactic medium (IGM) through galactic winds is accounted for as well, in terms of mass/energy exchange from all star forming galaxies and surrounding medium. Once the stellar mass is known, its metallicity and age are used to derive the spectra and emissivity of each galaxy and to compute the radiative transfer (RT) of the escaping radiation through the cosmic volume. The RT also derives the in-homogeneous gas ionization and temperature patterns in the IGM; this is finally used to determine the star formation process in the successive generations of structures, by suppressing the accretion of new gas mass in the galaxies\footnote{Note that the RT module can be switched off in favour of a more simple and analytic feedback scheme accounting for instant reionization at fixed redshift. This configuration as assumed in this work.}. 
To follow the redshift evolution down to the local universe, the above algorithm is applied at each snapshot $i$ by deriving the initial conditions from the baryonic quantities computed at $i-1$: these are transferred from ancestors to descendants by relying on a particle-based halo merger tree and consistently with the dynamical processes regulating the mass transfer, as described above (we refer to the original publications for more technical details). 

The \texttt{GAMESH} pipeline is then capable to account for mechanical, chemical and radiative feedback during the formation and evolution of the galaxies evolving in the cosmic volume fixed by the DM simulation. Note, on the other hand, that an hydrodynamical description of the gas dynamics is missing in our model because the mass budget relies only in the scaling with dynamical interactions between DM particles (i.e. DM halo collapse, mergers, tidal stripping and halo destruction). 

While the \texttt{GAMESH} model can run in principle on any DM simulation, the latest implementation \citep{2017MNRAS.469.1101G} relies on a multi-mass-resolution, zoom-in simulation of a highly resolved cosmic volume ($\sim4$~cMpc/side) containing a MW galaxy forming at its center and a plethora of companions galaxies. At $z=0$ we find a total collapsed mass of $M_{\texttt{DM}} \sim 3 \times 10^{12} M_{\odot}$ distributed in $2458$ halos; the MW-like halo dominates with a DM mass $M_{\texttt{MW}} \sim 1.7 \times 10^{12} M_{\odot}$, while many companion halos are found in different mass ranges: two have a DM mass of $M \sim 10^{11}$~M$_{\odot}$ (similar to the one observed in M32 or M33), while 14 halos have a mass in $10^{10} \lesssim M < 10^{11}$~M$_{\odot}$; finally a third population of 98 intermediate-DM halos are distributed in the mass range $10^{9} \lesssim M < 10^{10}$~M$_{\odot}$. A plethora of mini-halo satellites is also found in the entire volume, with N$_{\rm Mini} \sim 550$ orbiting within $2\times R_{\rm vir}$ from the central MW. The global statistic of objects in described in Figure 4 of the paper mentioned above. Finally note that this simulation is not meant to reproduce the dynamical properties of the observed Local Group but zooms-in on a single MW-like halo and its environment. Large scale DM simulations predict a statistical distribution of MW-like galaxies both in isolated and binary configurations; finding the  ICs leading to a final dynamical configuration comparable with the Local Volume is still and active research field (see for e.g. the CLUES collaboration  \citet{2017MNRAS.465.4886C}). For example, the simulation adopted in this work also predicts an Andromeda-like companion of the central MW, but this is found at a distance higher than 2~cMpc and then contaminated by low-mass resolution particles. The present simulation cannot then be used to  investigate the effects of MW-M31 dynamical co-evolution \citep{2014MNRAS.438.2578G} of for a direct dynamical comparison with the observed dynamics of the Local Group \citep{2015ApJ...807...49W}. 

Despite the above limitations\footnote{Note that new simulations based on the CLUES DM runs are under investigation.}, the accurate feedback model implemented in \texttt{GAMESH} has been proven to reproduce the baryonic properties of the MW galaxy and many scaling relations observed in our Local Universe. \citet{2017MNRAS.469.1101G} have shown for example that with a unique set of efficiency parameters regulating star formation and feedback efficiency\footnote{We calibrate these parameters by requiring the star formation rate, the stellar and gas masses, and the metallicity of the simulated MW galaxy at $z =0$ to match the observationally-inferred values of \citet{2006MNRAS.372.1149F}, \citet{2011MNRAS.414.2446M}, \citet{2015ApJ...806...96L}, \citet{2013ApJ...779..115B}, \citet{2015A&A...580A.126K} for stellar and gas masses. The mass of metals at $z=0$ is constrained by the values in \citet{2014ApJ...786...54P}} it is possible to reproduce a plausible SF history of a MW-like galaxy:

\begin{itemize}

\item at $z=0$ the \texttt{GAMESH} predictions are consistent with values observed in the Milky Way for both stellar, metal and gas mass as well as for its star formation rate (SFR, see Figure 5 and 6 in the original paper).  Their redshift evolution is also compatible with predictions from pure SAM models based on semi-analytic merger trees (based on the extended Press-Schechter formalism) and alternative models with a similar SF history ( e.g. model $m_3$ of \citet{2016A&A...589A.109C});

\item in the redshift range $0 < z < 4$ the SFR, stellar, gas and metal masses of the MW progenitor were compared with recent studies (e.g. \citet{2015ApJ...803...26P}, see Figures 7-8 in the reference paper) finding that when the mass selection includes systems at $z > 1.5$, they follow a stellar mass trend in good agreement with the observations, 
although their gas fraction have a shallower evolution in the 3 Gyr period between $z = 2.5$ and $z = 1$\footnote{This discrepancy might be induced by the assumed Instantaneous Recycling Approximation, which affects the time evolution of individual galaxies or by the lack of in-homogeneous radiative transfer effects.}. Also note that the more massive MW progenitor at each redshift lies within only a factor of 2 of the galaxy main sequence all the way from $z \sim  2.5$ to $z \sim 0$; 

\item we find (see Figure 9-10 same paper) that the distribution of the most massive MW progenitors is consistent with the fundamental plane of metallicity \citep[FPZ]{2012MNRAS.427..906H,2016MNRAS.463.2002H} and aligned with the fundamental metallicity relation \citep[FMR]{2010MNRAS.408.2115M}. These agreements with redshift-independent scaling relations involving metallicity, stellar
mass and star formation rate lets us to conclude conclude that the the feedback processes implemented in \texttt{GAMESH} leads to results consistent with observations even in a simulated, biased region of the current Universe.

\end{itemize}  

Further investigations on the chemical properties of the galaxies in the \texttt{GAMESH} volume can be found in \citet{2018MNRAS.473.4538G}, after introducing dust production in the model. The chemical properties of these galaxies are compared with the DGS and KINGFISH \citep{2011PASP..123.1347K} catalogs. Figure 2 and 3 in this study clearly show that the observed sample of low-stellar and low-metallicity galaxies is well covered by our simulation. 

As in the present work we extend the DGS sample with the ALLSMOG catalog and search for dwarfs hosting GW150914-like events, here we provide additional details on the statistic of star forming objects at $z=0$ in our simulation\footnote{Note that, with this constraint, only Lyman-$\alpha$-cooling halos are selected as they are the only objects with SFR($z=0)$ >0 in our simulation, and then can have an observable counterpart.}. In this way the reader can understand if the galaxies we found in Section \ref{sec:sample} are peculiar or simply common, representative objects\footnote{An extended comparison of the entire dwarf sample will be the topic of a future work exploiting the RT capabilities of our pipeline.}. In Figure \ref{fig:GAMESH_DWARFS} we provide the statistical distribution of these halos in terms of stellar mass (left panel) and metallicity (right panel). The blue histograms represent simulated data while the range of values found in the  DGS and ALLSMOG are shown as green and violet areas, respectively. The top panels of the same figure show the normalized histograms of the observed samples with same colour conventions.
 
The stellar mass of the \texttt{GAMESH} galaxies largely covers the low-end of the observed sample ($log(M_{\star}/M_{\odot}) > 6.5$), while our small volume induces an expected, limited coverage of the high-mass population of the ALLSMOG sample (note the gap in the simulates sample between ($9.8 < log(M_{\star}/M_{\odot}) < 10.2)$). 

\begin{flushleft}
\begin{figure}
\hskip -0.7truecm
\includegraphics [clip=true, width=9.50cm]{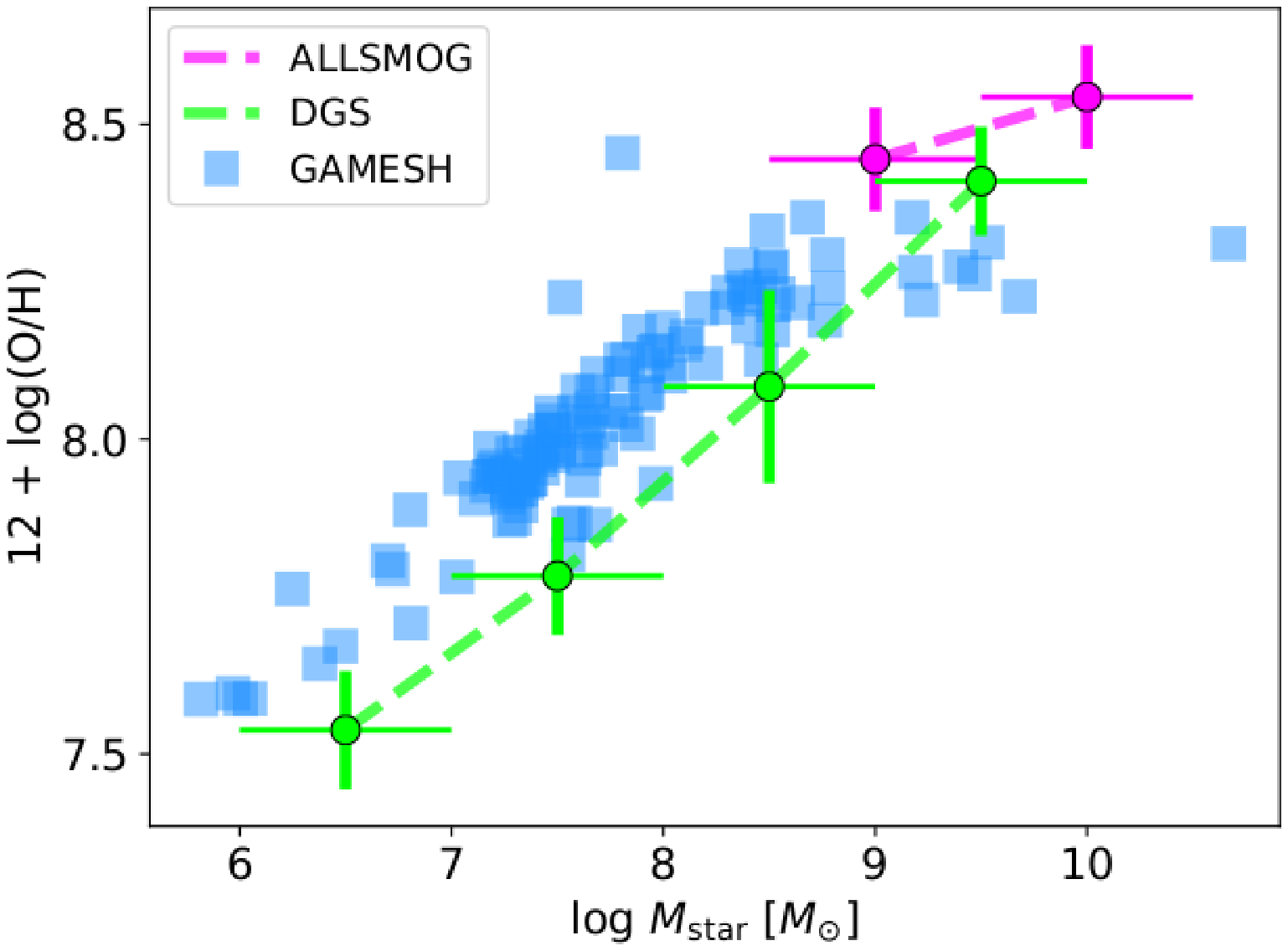}
\caption{ The mass metallicity relation at $z=0$ of both observed (violet/green)
and simulated data  (blue squares). The trend of the median values in the DGS survey are represented as green
lines, while violet lines refer to the ALLSMOG. The scatter
in the data can be inferred by the size of the error bars. The
simulated galaxies are represented instead as blue squares
and show that the GAMESH galaxies have a trend in good
agreement with the observed relation. 
}
\label{fig:MZRelation}
\end{figure}
\end{flushleft}

Our sample does not show any star forming galaxy with $Z < 7.75$ while it entirely covers the rest of the DGS and the low stellar mass tail of the ALLSMOG survey. Low metallicity, star-forming objects are then missing at $z=0$ in our simulation and this could derive from the  absence of important hydrodynamical details in predicting the efficiency of galactic winds in these extreme dwarfs. Other feedback effects could also have some role in establishing the configuration of these galaxies at $z=0$: the IRA approximation adopted in run can certainly induce a systematical early enrichment of all the galaxies because it does not account for the time dependence in their stellar evolution. We defer these points to future investigations when the full feedback model of \texttt{GAMESH} will be exploited in the current high-resolution DM simulation and the relative contribution of all the approximations mentioned above could be better established. Finally note that the remaining range of observed metallicity is well covered by our sample, with the usual bias in high mass objects due to the small volume of our simulation.

The mass metallicity relation at $z=0$ of both observed and simulated data is shown in Figure \ref{fig:MZRelation}. The  trend of the median values in the DGS survey are represented as green lines, while violet lines refer to the ALLSMOG. The scatter in the data can be inferred by the size of the error bars. The simulated galaxies are represented instead as blue squares and show that the \texttt{GAMESH} galaxies have a trend in good agreement with the observed relation (also see the original paper for a comparison with \citealt{2010MNRAS.408.2115M}) while the simulated metallicity seem systematically higher than the observed one. It should be noted, on the other hand, that the DGS sample is well known to lie slightly below the M-Z, due to a combination of metallicity calibration and high star formation rate efficiencies; starburst galaxies in fact, usually exhibit lower $Z$ at fixed M$_{\star}$ (see for example \citealt{2010MNRAS.408.2115M,2016A&A...595A..48B}).
   
\section{The SeBa database}
\label{sec:SeBa_DB} 
In this study we adopt the BPS code \texttt{SeBa}, recently modified by \citet{2013MNRAS.429.2298M} to explicitly account for a metallicity-depend evolution of the stars, of their stellar winds and the formation of their remnants; the other aspects of the binary evolution are instead unchanged with respect to the original version developed by \citealt{1996A&A...309..179P} and \citealt{2001A&A...375..890N}.
In our \texttt{GAMESH} extension, \texttt{SeBa} has been used to generate many databases of binary systems spanning different ranges of configurations by mimicking the strategy described in \citet{2017MNRAS.471L.105S}, where the evolution of a large number of binary systems has been simulated in 11 different values of the star metallicity, covering the  range $0.01 \leq Z/Z_\odot \leq 1 $. Their initial properties were randomly selected from independent distribution functions. In particular: 

\begin{itemize}

\item the primary stellar mass, $m_{\rm p}$, is distributed according to a Kroupa Initial Mass Function (IMF, \citealt{2001MNRAS.322..231K}) sampled in a suitable mass range, while the mass of the secondary star, $m_{\rm s}$, is generated according to a flat distribution for the mass ratio $q = m_{\rm s}/m_{\rm p}$ with $0.1 < q \leq 1$;

\item the initial semi-major axis (sma) distribution is flat in log(sma) (see \citealt{1996A&A...309..179P}), 
ranging from $0.1 \,R_{\odot}$ (Roche lobe contact), up to $10^{6} \, R_{\odot}$;

\item the eccentricity of the binary is selected from a thermal distribution $f(e) = 2e$ in the $[0,1]$ range \citep{1975MNRAS.173..729H}.

\end{itemize}
 
To obtain a significant statistic of the most massive candidates, here, a series of new databases has been generated in the IMF mass range $[8.0,100]$~M$_\odot$, while keeping the other initial conditions. Table \ref{tab:numSystems} summarizes the properties of the various databases described in this section.  In DB$_{\texttt{ID}}=1$ N$_b= 2 \times 10^{6}$ systems are appropriate to provide a variegate statistic of the evolution of high mass component binary systems at low metallicity. The resulting  number of candidates potentially generating the GW150914-like signals is $\approx 290$ in the three bins shown in table, while drop to zero at $Z > 0.05$~Z$_\odot$. A second database of N$_b= 2 \times 10^{7}$ samples the IMF in $[0.1, 100.0]$~M$_\odot$ and provides only 10 suitable systems, indicating sampling a wider mass range is not sufficient to capture the statistic of high-mass binaries. Finally, DB$_{\texttt{ID}}$ = 3 finds more than 2900 massive systems, providing the right statistic in which higher mass components could be found as suggested by \citet{2018arXiv181112907T}, and will be useful for further investigations.

\begin{table}
\begin{centering}
\begin{tabular}{|c|c|c|c|c}
\hline
DB$_{\texttt{ID}}$ & N$_b$ & IMF mass range [M$_\odot$] & Z~[Z$_\odot$] & N$_{\texttt{GW150914}}$ \tabularnewline
\hline
\hline
1 & $2 \times 10^{6}$ & $[8.0, 100]$ & 0.01 & $\approx290$\tabularnewline
\hline
1 & $2 \times 10^{6}$ & $[8.0, 100]$ & 0.02 & $\approx290$\tabularnewline
\hline
1 & $2 \times 10^{6}$ & $[8.0, 100]$ & 0.05 & $\approx290$\tabularnewline
\hline
1 & $2 \times 10^{6}$ & $[8.0, 100]$ & 0.06 & $0$\tabularnewline
\hline
2 & $2 \times 10^{7}$ & $[0.1, 100]$ & 0.01 & $\approx10$\tabularnewline
\hline
3 & $2 \times 10^{7}$ & $[8.0, 100]$ & 0.01 & $\approx2900$\tabularnewline
\hline
\end{tabular}
\par\end{centering}
\caption{\label{tab:numSystems} \texttt{SeBa} databases (DB$_{\texttt{ID}}$) generated at different stellar metallicity ($Z$), with a different total number of simulated binaries (N$_b$) and in different ranges of  IMF mass. Note that binary black holes compatible with the mass estimates for the GW150914 signal (N$_{\texttt{GW150914}}$) are not zero only below $Z=0.05$~Z$_\odot$ and that only with a suitable choice of the IMF mass range, $N_b$ becomes statistically significant. }
\end{table}

Before concluding this section we point out that the generation of each randomly sampled \texttt{SeBa} database (at fixed metallicity) has its internal random noise depending on the specific choice of initial parameters. Its statistical representativeness must then be tested across different realizations relying on different chains of random numbers.    

\begin{table}
\begin{centering}
\begin{tabular}{|c|c|c|c}
\hline
DB$_{\texttt{ID}}$ & RNG Seed ID & Z~[Z$_\odot$] & N$_{\texttt{GW150914}}$ \tabularnewline
\hline
\hline
1 & 0 & 0.01 & 287\tabularnewline
\hline
1 & 1 & 0.01 & 270\tabularnewline
\hline
1 & 2 & 0.01 & 294\tabularnewline
\hline
1 & 3 & 0.01 & 297\tabularnewline
\hline
1 & 4 & 0.01 & 327\tabularnewline
\hline
1 & 0 & 0.05 & 300\tabularnewline
\hline
1 & 1 & 0.05 & 306\tabularnewline
\hline
1 & 2 & 0.05 & 299\tabularnewline
\hline
1 & 3 & 0.05 & 290\tabularnewline
\hline
1 & 4 & 0.05 & 282\tabularnewline
\hline
2 & 0 & 0.01 & 14\tabularnewline
\hline
2 & 1 & 0.01 & 5\tabularnewline
\hline
2 & 2 & 0.01 & 6\tabularnewline
\hline
2 & 3 & 0.01 & 11\tabularnewline
\hline
2 & 4 & 0.01 & 15\tabularnewline
\hline
\end{tabular}
\par\end{centering}
\caption{\label{tab:SystemConvergence} \texttt{SeBa} databases convergence in number of predictions for binaries compatible in mass with GW150914 event (N$_{\texttt{GW150914}}$). The Database ID (DB ID) refers to the one in Table \ref{tab:numSystems} and the seed unique ID used to extract a random chain of the RNG adopted by \texttt{SeBa} is indicated with RNG seed ID and it is consistent across realizations.  The convergence in N$_{GW150914}$ is shown for the metallicity bins [0.01, 0.05]~Z$_\odot$ of DB$_{\texttt{ID}}$ = 1 and for Z=0.01~Z$_\odot$ of DB$_{\texttt{ID}}$=2.}

\end{table}

Table \ref{tab:SystemConvergence} shows the results of convergence tests performed generating identical databases with different chains of random numbers at assigned random number generator (RNG) seed. For example, we show that 5 realizations of 2 metallicity bins of DB$_{\texttt{ID}}$ = 1 provide quite stable number of predicted high-mass binaries ($\approx 290$) because of the large number of sampling systems available in the mas range [8.0, 100]~M$_\odot$. In the case of DB$_{\texttt{ID}}$ = 2, spanning a wider range of stellar masses, the number of predicted systems remains instead too low, as they are found between 5 and 15. 

\section{Selection of binary systems in GAMESH}

The last implementation of \texttt{GAMESH} combines the model described in Appendix \ref{sec:GAMESH_MODEL} with the \texttt{SeBa} database (Appendix \ref{sec:SeBa_DB}) to study the formation and coalescence sites of the GW events detected by the LIGO Collaboration. This extension works by identifying binary systems in star forming halos and by following their evolution until coalescence; this is done self-consistently with the cosmological evolution described in the \texttt{GAMESH} volume.

In this section we first describe the above method, second we study the statistical convergence of its predictions.

In each star forming halo $j$, found at a certain redshift $z_i$ (SFR$_{j,i}>0$), the algorithm determines the newly formed stellar mass in binaries as ${M_{2 \star}}^{j,i} = f_{2\star}\times{M_{\star}}^{j,i}$, where $f_{2\star}$ is the redshift-indepedent binary fraction. It then samples it in the metallicity bin $k$ of \texttt{SeBa} by requiring the gas metallicity ($Z_{j,i}$) to be  $Z_{k-1}<Z_{j,i}<Z_{k+1}$. This is done in two steps: first $Z_{j,i}$ is used to identify the right database sub-sample, and second, an appropriate number of binary systems is randomly selected in $k$ until the sum of their component masses exhausts ${M_{2 \star}}^{j,i}$. 

The result of this process associates all the selected binary system IDs of \texttt{SeBa} with the ID of the  \texttt{GAMESH} halos. Each binary is then assumed to form in the halo $j$ at $z_i$, and its evolution can be unambiguously followed along the halo accretion history through the particle-based merger tree of \texttt{GAMESH}\footnote{Note that the stellar mass is followed along the merger tree by scaling it by mass transfer ratios from ancestors contribution and by assuming the stellar mass to be located at the center of the DM halos. As the simulation is dominated by gas accretion and minor mergers (see the DM simulation description in \citet{2017MNRAS.469.1101G}), these two assumptions guarantee that there is no ambiguity in the transfer of GW150914-like systems along this process from an ancestor to the main descendant.}. For example, at $z_{i+1}$ the evolutionary status of each binary system is established by checking the status of its components after a physical time $\Delta t$ separating the snapshots $z_{i+1}$ and $z_i$. In this way the correct evolutionary status of each binary is described in the descendant halo at $z_{i+i}$. By repeating the above algorithm along the cosmological simulation we are capable to track the evolution of all the formed binaries at all redshifts. 

While the algorithm described above is sufficiently general to follow the evolution of the entire population of binary systems, for specific applications investigating a precise class of binaries (as in the present paper), the computation can be significantly improved by sampling ${M_{2 \star}}^{j,i}$ in restricted intervals of the IMF. For example, to recognize massive binaries generating GW150914-like events, it is sufficient to restrict the IMF mass range to $[8.0,100.0]$~M$_\odot$ and, by assuming a SALPETER-like shape, to sample $\sim14$~\% of the contained ${M_{2 \star}}^{j,i}$. An accurate choice of the \texttt{SeBa} database is necessary to ensure the right statistical convergence of the results. In Appendix \ref{sec:SeBa_DB} we have shown that a database with N$_b= 2 \times 10^6$ systems sampled in $[8.0,100.0]$~M$_\odot$ (see e.g. DB$_{\texttt{ID}}$ = 1) provides a statistically convergent number of massive binaries in which GW150914 candidates can be found. Moreover, only dwarfs with $Z_{\star} < 0.05$ must be sampled to find massive binaries, as shown in Table \ref{tab:numSystems}.

We conclude this appendix by testing the statistical robustness of the dwarfs sampling discussed in the present work. In fact, the binary systems association procedure is implemented with a random sampling strategy and its convergence must be verified. For this reason, we ran the association in galaxy dwarfs with different chains of the RNG and tested the resulting formation sites on a series of randomly selected redshift snapshots. These results are summarized in Table \ref{tab:SysAssociationConvergence} for a case at $z\approx 4$ and two random realizations.

\begin{table}
\begin{centering}
\begin{tabular}{|c|c|c|c|c|c}
\hline
Run$_{\texttt{ID}}$ & RNG Seed ID & DB$_{\texttt{ID}}$ & N$_{\texttt{SeBa Ids}}$ & N$_{\texttt{fH}}$& N$_{\texttt{GW150914}}$  \tabularnewline
\hline
\hline
1 & 0 & 1 & 18 & 20 & 5 \tabularnewline
\hline
2 & 1 & 1 & 21 & 26 & 8 \tabularnewline
\hline
\end{tabular}
\par\end{centering}
\caption{\label{tab:SysAssociationConvergence} \texttt{GAMESH} runs associating star forming halos with new binaries at $z\approx4.0$. In each table row we show the unique run identifier (Run$_{\texttt{ID}}$), the unique RNG seed adopted for each run (RNG Seed ID), the identifier of the adopted \texttt{SeBa} database (DB$_{\texttt{ID}}$) as in Table \ref{tab:SystemConvergence}, the number of halos hosting their formation (N$_{\texttt{fH}}$) and finally the number of different binary systems corresponding to good candidates to generate GW150914 at coalescence (N$_{\texttt{GW150914}}$).}

\end{table}

The newly formed stellar mass was sampled from DB$_{\texttt{ID}}$ = 1, providing a large variety of binary channels with masses compatible with the predictions for GW150914. The number of formation halos (N$_{\texttt{fH}}$) involved in both realizations is compatible as they are 20 and 26 respectively, and all the dwarfs categories analyzed in this paper (E-systems, F-systems, MW-systems) are well represented in each obtained sample. Also note that they host a compatible number of binary candidates (N$_{\texttt{SeBa Ids}}$), i.e. 18 and 21. The last column (N$_{\texttt{GW150914}}$) shows that 5 and 8 of those events can be found as GW150914-like coalescing events as constrained by their coalescence time.
Note that to study the evolution of the dwarfs hosting these systems, we selected representative histories found in the entire sample and belonging to both random realizations.      

\end{appendix}

\label{lastpage}
\end{document}